\documentclass[twocolumn, prx, reprint, amsfonts, amsmath, amssymb, superscriptaddress, longbibliography, floatfix,aps,groupedaddress]{revtex4-1}
\usepackage{comment} 
\usepackage{bm}

\usepackage[T1]{fontenc}
\usepackage{amsmath,amssymb}
\usepackage{graphicx}
\usepackage[dvipsnames]{xcolor}
\usepackage{braket}
\usepackage{enumitem}
\usepackage[english]{babel}
\usepackage{titlesec} 
\usepackage[colorlinks,linkcolor=Blue,citecolor=Blue]{hyperref}
\usepackage{tikz}
\usepackage{circuitikz}
\usetikzlibrary{positioning}
\usepackage{multirow}
\usepackage{pifont}
\usepackage{booktabs}

\newcommand{\linpol}{\ensuremath{\updownarrow}}
\newcommand{\circpol}{\ensuremath{\circlearrowright}}

\newcommand{\yesmark}{\ding{52}}
\newcommand{\nomark}{\ding{56}}

\begin{document}
\title{Non-Hermitian Photocurrents and Skew Quantum Geometry}

\author{Yinning Niu} 
\affiliation{Division of Physics and Applied Physics, School of Physical and Mathematical Sciences, Nanyang Technological University, Singapore 637371}

\author{Xu Yang}
\email{xu.yang@ntu.edu.sg}
\affiliation{Division of Physics and Applied Physics, School of Physical and Mathematical Sciences, Nanyang Technological University, Singapore 637371}

\author{Justin C.W. Song}
\email{justinsong@ntu.edu.sg}
\affiliation{Division of Physics and Applied Physics, School of Physical and Mathematical Sciences, Nanyang Technological University, Singapore 637371}

\date{\today}

\begin{abstract}
    We argue that 
    indirect optical transitions have a {\it non-Hermitian} character that dramatically changes the nature of quantum geometric optical response. In particular, we introduce a set of non-Hermitian optical transition photocurrents with opposite properties to that of conventional bulk quantum geometric photocurrents. For example, we 
    find circular non-Hermitian shift photocurrents persist in non-magnetic systems where conventional circular shift photocurrents are known to {\it vanish}. Strikingly, we find such non-Hermitian circular shift photocurrents dominate the circular photogalvanic effect of excitons in non-magnetic  materials.  
    We map an expanded quantum geometry of resonant optical transitions  
    and find two essential classes: symmetric and skew. 
    The former describes reciprocal transitions and conventional bulk quantum geometric photocurrents. The latter captures antisymmetric transitions and non-Hermitian photocurrents. Taken together, these unify scattering and intrinsic photocurrents under a single non-Hermitian quantum geometric umbrella. 
    These demonstrate how scattering {\it enriches} quantum geometry and produces optoelectronic responses beyond the simple direct and Hermitian optical transitions. 
\end{abstract}

\maketitle

The electric dipole plays a central role in linear and nonlinear optics~\cite{vonBaltzKraut1981, SipeShkrebtii2000, MorimotoNagaosa2016, ChingKit2017, DeJuan, Parker, AhnGuoNagaosa2020, WatanabeYanase2021, AhnEtAl2022}: its amplitude determines transition strength~\cite{SipeShkrebtii2000, AhnEtAl2022} and its geometry~\cite{MorimotoNagaosa2016,AhnGuoNagaosa2020,WatanabeYanase2021,AhnEtAl2022,Parker,DeJuan} controls bulk quantum geometric nonlinearities such as photocurrent and photoconductivity~\cite{AhnEtAl2022} responses. Because the electric dipole is Hermitian, direct electric dipole optical transitions and their associated responses are strongly constrained. A prime example are shift photocurrents arising from real space shifts an electron undergoes under light irradiation~\cite{vonBaltzKraut1981, SipeShkrebtii2000, MorimotoNagaosa2016}. The Hermitian character of the direct electric dipole transitions zeros circular shift photocurrents in non-magnetic materials~\cite{AhnGuoNagaosa2020, WatanabeYanase2021}; linear shift photocurrents, similarly, vanish in space-time reversal symmetric magnets~\cite{AhnGuoNagaosa2020, WatanabeYanase2021}. As a result, symmetry has been used as a guiding principle in delineating quantum geometric photocurrent~\cite{DongSun2019,Osterhoudt2019,XuEtAl2018, KangEtAl2025, MaEtAl2023,DongEtAl2026}.

\begin{figure}[t]
    \centering
    \includegraphics[scale=0.5]{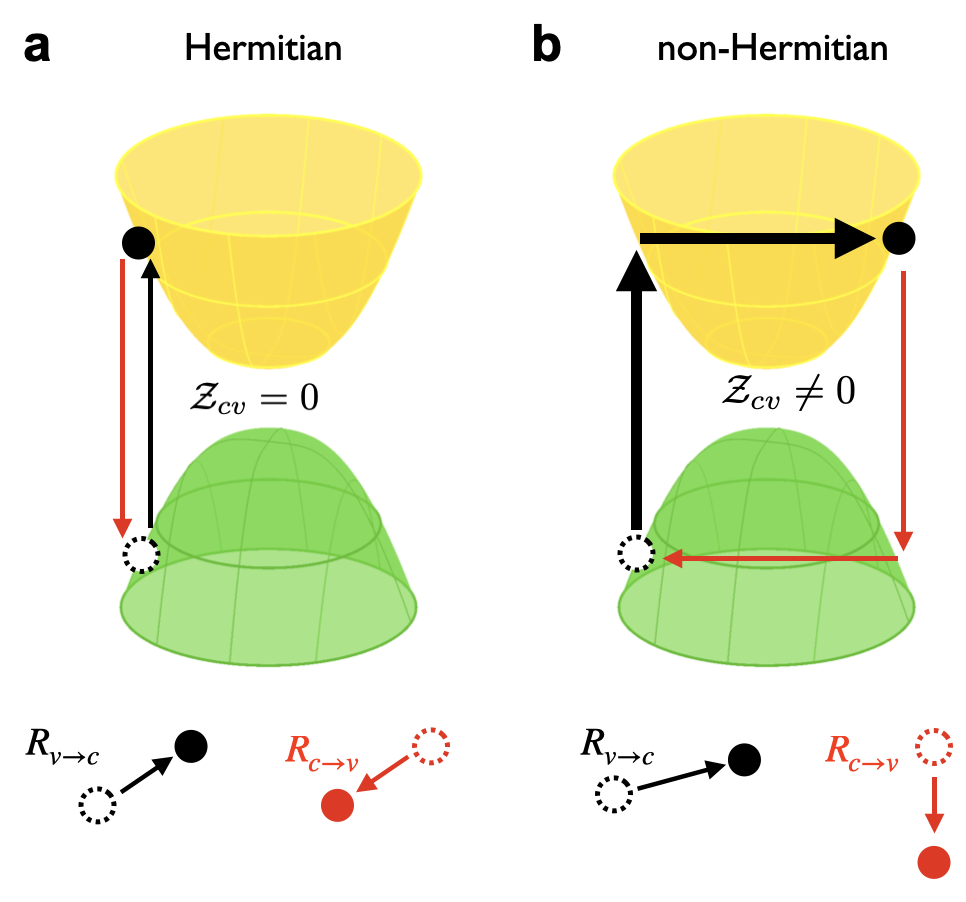}
    \caption{\textbf{Hermitian and non-Hermitian optical transitions.} Optical transitions are depicted by black and red arrows that denote forward ($v\to c$) and reverse ($c\to v$) optical transition processes. {\bf a.} Direct optical transitions have a transition rate that is the same for the forward and reverse process $W_{v\to c} = W_{c\to v}$. The shift vector for the forward and reverse transitions shift vectors have the same magnitude but opposite directions. These two properties arise from its Hermitian character. {\bf b.} In contrast, indirect optical transitions can have a non-Hermitian character with unequal forward and reverse rates $W_{v\to c} \neq W_{c\to v}$, and uncompensated shift vectors for forward and reverse processes $\mathbf{R}_{v\to c} \neq - \mathbf{R}_{c\to v}$. This unconventional behavior arises from a non-Hermitian interference Wilson-loop $\mathcal{Z}_{cv}$ that controls both the unequal transition rates as well as the uncompensated geometric shift vectors. Note that Hermitian optical transitions have $\mathcal{Z}_{cv} = 0$. The top panels depict transition rates while bottom panels track the shift vector for each of the forward (black) and reverse (red) processes. }
    \label{Fig 1}
\end{figure}

Unlike direct optical transitions that have a Hermitian character [Fig.~~\ref{Fig 1}a], indirect transitions are different. Instead, phase accumulation during multiple scattering can render such indirect transitions non-Hermitian~\cite{Sinitsyn2008, NagaosaEtAl2010, McGinleyCooper2020}. This non-Hermiticity arises even as the light-matter interaction and scattering potentials are by themselves Hermitian. 
Such multiple impurity scattering can qualitatively change electronic response: non-Hermitian $T$-matrices underpin skew scattering important for extrinsic anomalous Hall effects \cite{Sinitsyn2008, NagaosaEtAl2010} and induce fragility in topological phases protected by anti-unitary symmetries \cite{McGinleyCooper2020}. 
In these examples, non-Hermiticity affect transition rates and produce unconventional dynamics. 

Here we argue that non-Hermitian optical transitions [Fig.~\ref{Fig 1}b] processes produce anomalous geometric phases 
with dramatic effects on opto-electronic response. In particular, we introduce a class of non-Hermitian photocurrents: 
non-Hermitian shift photocurrents. These non-Hermitian shift photocurrents contrasts sharply with their Hermitian counterparts. For example, we find non-Hermitian circular shift photocurrents survive in nonmagnetic materials (i.e. time-reversal $\mathcal{T}$-symmetric materials);  linear polarized non-Hermitian shift photocurrents survive in space-time-reversal $\mathcal{PT}$-symmetric magnets. The conventional (Hermitian) shift currents from direct electric dipole transitions vanish in both cases. This indicates scattering enables a richer class of photocurrents than was previously thought possible and distinct from the conventional quantum geometric photocurrent symmetry classification~\cite{WatanabeYanase2021, AhnGuoNagaosa2020}. 

At a fundamental level, we trace this unconventional behavior to how optical transitions can be described by a non-Hermitian quantum geometry that is characterized by a non-Hermitian interference Wilson-loop. The conventional (Hermitian) quantum geometric description of optical response~\cite{AhnEtAl2022, AhnGuoNagaosa2020}
is based on the direct transitions driven by the electric dipole. This produces a geometry that is identical (up to a complex conjugate) for optical transitions from valence ($v$) to conduction ($c$) bands and transitions from $c \to v$ bands.  
In contrast, we find non-Hermitian processes break this simple $c\leftrightarrow v$ reciprocity and, instead, possess a finer structure comprising a symmetric quantum geometry and a skew quantum geometry. As we explain below the symmetric quantum geometry track Hermitian transitions Fig.~\ref{Fig 1}a while the skew quantum geometry describe non-Hermitian transitions Fig.~\ref{Fig 1}b. Indeed, we find the skew quantum geometry are directly related to optoelectronic responses: the imaginary (real) parts of the skew quantum connections produce the linear (circular) non-Hermitian shift photocurrent we introduce here.

Interestingly, the real part of the skew quantum metric tracks the Ballistic current \cite{BelinicherSturman1988, BelinicherSturman1980, LeppenenGolub2023, Sturman2020, SturmanFridkin1992, DaiEtAl2021, Ivchenko, DaiRappe2021} (linearly polarized light induced injection type photocurrents even in a nonmagnetic system). Such Ballistic currents have conspicuously fallen outside of the regular classification of quantum geometric photocurrents within simple (Hermitian) electric-dipole treatments. Instead, Ballistic currents are naturally produced by non-Hermitian optical processes, and as we argue, can be readily described in terms of a skew quantum geometry. Additionally, we find the imaginary part of the skew quantum metric is no less important: it produces a $\mathcal{T}$-odd circular non-Hermitian injection current that persists even in $\mathcal{PT}$-symmetric systems; the conventional circular injection current vanishes in such space-time magnets. Together with the skew quantum connections, these provide a rich description of non-Hermitian interband optical processes. 

 We argue non-Hermitian shift photocurrents are especially important for excitonic transitions where injection (and ballistic) photocurrents {\it vanish} due to the bound localized nature of the particle-hole pairs \cite{Wannier1937, YangSrivastavaSong2026}.
Instead, shift photocurrents are expected to dominate the photocurrent generation for excitonic transitions \cite{YangSrivastavaSong2026, DongEtAl2023, AkamatsuEtAl2021}. We find that the non-Hermitian circular shift photocurrent dominates the excitonic transition photocurrent in non-magnetic materials enabling direct access to the skew quantum connection and dominates the circular photogalvanic effect for excitons in non-magnetic materials. Similarly, even as discrete rotational symmetry zeroes vertical transition excitonic shift photocurrents~\cite{YangSrivastavaSong2026}, we argue that disorder assisted processes nevertheless enable linear excitonic transition shift photocurrents. Such indirect excitonic transition shift currents that we discuss closely mirror recent unexplained shift current experimental measurements in excitonic system 3R-MoS$_2$~\cite{DongEtAl2023}. 

Our work highlights the unconventional role of non-Hermiticity in {\it interband} optical transitions and how it affects both responses (e.g., interband photocurent) as well as the resonant quantum geometry. This dramatically expands the means wherein disorder processes affect optical response \cite{Ishizuka2021,Kaplan2020}; non-Hermitian optical transitions, as we argue, change symmetry properties for optical transitions, unblock photocurrents, and provide a richer set of geometric phenomena not previously known. Taken together, these demonstrate the pervasive and necessary role of an expanded non-Hermitian quantum geometry of indirect optical transitions in tracking optical response. 

\section{Non-Hermitian optical transitions} 

We begin by examining the rate of optical excitation of electronic states in the presence of electromagnetic radiation and disorder. Here we will consider a total potential $\hat V(t)=\hat V_{\rm dis}+\hat V_\omega e^{-i\omega t}+\hat V_\omega^\dagger e^{i\omega t}$ where $\hat{V}_{\rm dis}$ is a disorder potential and $\hat V_\omega=-e\hat{\mathbf r}\cdot\mathbf E(\omega)$ is the light-matter interaction for an oscillating electromagnetic field at frequency $\omega$. The optical transition rate from $|\alpha\rangle$ to $|\beta\rangle$ can be expressed within a $T$-matrix formulation \cite{NagaosaEtAl2010,Sinitsyn2008} as $W_{\alpha\to\beta}=\sum_{\omega}W_{\alpha\to\beta}^{(\omega)}$, where the sum includes both absorption and emission processes with
\begin{equation}
 W_{\alpha\to\beta}^{(\omega)}
 =\frac{2\pi}{\hbar}\Delta_\omega
 |T_{\beta\alpha}(\omega)|^2,
 \quad
 \Delta_\omega
 =\delta(\epsilon_\beta-\epsilon_\alpha-\hbar\omega),
 \label{eq:rate}
\end{equation}
where $T_{\beta\alpha}(\omega)=\langle\beta|\hat T(\omega)|\alpha\rangle$ is the transition amplitude for the frequency channel $\omega$. In the perturbative expansion
$\hat T=\hat{\mathbf V}+\hat{\mathbf V}\hat G_\eta^{\rm ret}\hat{\mathbf V}+\hat{\mathbf V}\hat G_\eta^{\rm ret}\hat{\mathbf V}\hat G_\eta^{\rm ret}\hat{\mathbf V}+\cdots$, and we retain, for $\hat T(\omega)$, exactly one $\hat V_\omega$ in each term, since we focus on the rectified photocurrent induced at second order in the applied electromagnetic field. Accordingly, we write $\hat T(\omega)=-e\sum_a\hat t^a(\omega)E^a(\omega)$, where $\hat t^a(\omega)$ is the field-free $T$-matrix operator for polarization component $a=x,y,z$, and $E^a(-\omega)=[E^a(\omega)]^*$ is the complex electromagnetic field. Here, $\hat G_\eta^{\rm ret}=(E-\hat H_0+i\eta)^{-1}$ is the retarded Green's function, with $E$ denoting the energy entering the corresponding propagator, and $\hat H_0$ is the bare electronic Hamiltonian satisfying $\hat H_0|\alpha\rangle=\epsilon_\alpha|\alpha\rangle$; here $\eta \to 0^+$ is an adiabatic parameter. A detailed derivation of the generalized $T$-matrix for optical transitions is presented in the Supplementary Material ({\bf SM}).

Critically, multiple scattering can render the field-free transition operators non-Hermitian, $\hat t^a\ne(\hat t^a)^\dagger$ \cite{McGinleyCooper2020}. Non-Hermitian operators can exhibit pronounced directionality; for example, giving distinct left and right eigenvectors \cite{ShenZhenFu2018}. For optical transitions, as we will see below, such non-Hermiticity can transform the optoelectronic response. To analyze it, we write $\hat t^a(\omega)=\hat{\mathcal X}^a(\omega)+i\hat{\mathcal Y}^a(\omega)$, with $\hat{\mathcal X}^a(\omega)=(\hat t^a(\omega)+[\hat t^a(-\omega)]^\dagger)/2$ and $\hat{\mathcal Y}^a(\omega)=(\hat t^a(\omega)-[\hat t^a(-\omega)]^\dagger)/(2i)$~\cite{McGinleyCooper2020}.
Using these, the transition amplitude product reads 
\begin{equation}
t_{\beta\alpha}^{a,\omega}(t_{\beta\alpha}^{b,\omega})^*
 \hspace{-1mm}=
 \mathcal X_{\alpha\beta}^{b,-\omega}
 \mathcal X_{\beta\alpha}^{a,\omega}
 +\mathcal Y_{\alpha\beta}^{b,-\omega}
 \mathcal Y_{\beta\alpha}^{a,\omega}+i\big[
 \mathcal Z_{\alpha\beta}^{ab,\omega}
 \hspace{-1mm}-(\mathcal Z_{\alpha\beta}^{ba,\omega})^*
 \big],
\label{eq:fieldfree-transition-kernel}
\end{equation}
where $\mathcal X_{\beta\alpha}^{a,\omega}=\langle\beta|\hat{\mathcal X}^a(\omega)|\alpha\rangle$ and $\mathcal Y_{\beta\alpha}^{a,\omega}=\langle\beta|\hat{\mathcal Y}^a(\omega)|\alpha\rangle$. Meanwhile, the cross-interference term reads $\mathcal Z_{\alpha\beta}^{ab}(\omega)\equiv
 \mathcal X_{\alpha\beta}^{b,-\omega}
 \mathcal Y_{\beta\alpha}^{a,\omega}$
\begin{equation}
\begin{aligned}[b]
 \mathcal Z_{\alpha\beta}^{ab}(\omega)=-\frac{i}{4}
 \langle\alpha|
 \hat t^b_{-\omega}+(\hat t^b_{\omega})^\dagger
 |\beta\rangle
 \langle\beta|
 \hat t^a_{\omega}-(\hat t^a_{-\omega})^\dagger
 |\alpha\rangle.
\end{aligned}
\label{NH-WL}
\end{equation}
We term $\mathcal Z_{\alpha\beta}^{ab}$ the non-Hermitian interference Wilson-loop tensor. It vanishes when the field-free transition operators are Hermitian, whereas the combination $\mathcal X_{\alpha\beta}^{b,-\omega}\mathcal X_{\beta\alpha}^{a,\omega}+\mathcal Y_{\alpha\beta}^{b,-\omega}\mathcal Y_{\beta\alpha}^{a,\omega}$ survives.
As we will argue throughout this article, $\mathcal Z_{\alpha\beta}^{ab}(\omega)$ governs non-Hermitian optical-transition phenomena ranging from nonreciprocal transition rates to anomalous shift vectors and skew quantum geometry. It reflects the interplay between geometric phase accumulation, scattering, and optical excitation, as discussed below.

The most direct consequence of $\mathcal Z_{\alpha\beta}^{ab}(\omega)$ is non-reciprocal optical transition rates $W_{\alpha\to\beta}\ne W_{\beta\to\alpha}$ that violate detailed balance \cite{BelinicherSturman1980,NagaosaEtAl2010}. In what follows, it will be useful to separate the linear and circular polarization contributions as
\begin{equation}
    E^a(\omega)E^b(-\omega)=L^{ab}(\omega)+iH^{ab}(\omega)
\label{eq:electric bilinear detachment}
\end{equation}
where $L^{ab}(\omega)=\operatorname{Re}[E^a(\omega)E^b(-\omega)]$ is real and symmetric, while $H^{ab}(\omega)=\operatorname{Im}[E^a(\omega)E^b(-\omega)]$ is real but antisymmetric. The $\mathcal Z_{\alpha\beta}^{ab}$ induced non-reciprocity is most readily captured by the antisymmetric transition rate $W_{\alpha\beta}^A=(W_{\alpha\to\beta}-W_{\beta\to\alpha})/2 = \sum_{\pm\omega}W_{\alpha\beta}^{A,(\omega)}$, so that
\begin{equation}
 W_{\alpha\beta}^{A,(\omega)}
 =-\rho_\omega\Big[
 L^{ab}(\omega)\operatorname{Im}\mathcal Z_{\alpha\beta}^{ab}(\omega)+H^{ab}(\omega)\operatorname{Re}\mathcal Z_{\alpha\beta}^{ab}(\omega)
 \Big],
\label{eq:antirate}
\end{equation}
where $\rho_\omega=4\pi e^2\Delta_\omega/\hbar$. Hereafter, we adopt the Einstein summation convention for repeated polarization indices $a,b$. This demonstrates that $W_{\alpha\beta}^A$ directly depends on the non-Hermitian interference Wilson-loop tensor $\mathcal Z_{\alpha\beta}^{ab}$ and vanishes when the field-free operators $\hat t^a$ are Hermitian [e.g., in the clean limit when $\hat t^a=\hat r^a$]. In contrast, the symmetric part of the transition rate reads $W_{\alpha\beta}^S=(W_{\alpha\to\beta}+W_{\beta\to\alpha})/2 = \sum_{\pm\omega}W_{\alpha\beta}^{S,(\omega)}$ where $W_{\alpha\beta}^{S,(\omega)}=\frac{\rho_\omega}{2}\Big\{L^{ab}(\omega)\operatorname{Re}\!\left[\mathcal X_{\alpha\beta}^{b,-\omega}\mathcal X_{\beta\alpha}^{a,\omega}+\mathcal Y_{\alpha\beta}^{b,-\omega}\mathcal Y_{\beta\alpha}^{a,\omega}\right]-H^{ab}(\omega)\operatorname{Im}\!\left[\mathcal X_{\alpha\beta}^{b,-\omega}\mathcal X_{\beta\alpha}^{a,\omega}+\mathcal Y_{\alpha\beta}^{b,-\omega}\mathcal Y_{\beta\alpha}^{a,\omega}\right]\Big\}$, which survives even when the field-free operators $\hat t^a$ are Hermitian (e.g., $\hat t^a=\hat r^a$). Notice that the antisymmetric $W_{\alpha\beta}^A$ directly corresponds to skew scattering \cite{BelinicherSturman1988,BelinicherSturman1980,LeppenenGolub2023,Sturman2020,SturmanFridkin1992,DaiEtAl2021, Ivchenko, NagaosaEtAl2010} obtained in a Born series, which is important for the anomalous Hall effect \cite{NagaosaEtAl2010}. Eq.~\eqref{eq:antirate} demonstrates that such antisymmetric rates generically proceed from non-Hermitian transitions characterized by the non-Hermitian interference Wilson-loop tensor $\mathcal Z_{\alpha\beta}^{ab}$.

For optical transitions, the non-Hermitian nature of $\mathcal Z_{\alpha\beta}^{ab}$ produces dramatic consequences for their symmetry properties. To see this, we examine anti-unitary symmetries with $\Theta$ an anti-unitary transformation (e.g., $\mathcal T$ or $\mathcal{PT}$) that can have particularly striking consequences for non-Hermitian transitions~\cite{McGinleyCooper2020}. We consider a bare Hamiltonian satisfying $\hat H_0|\alpha\rangle=\epsilon_\alpha|\alpha\rangle$ and $\Theta\hat H_0\Theta^{-1}=\hat H_0$, and scalar nonmagnetic disorder satisfying $\Theta\hat V_{\rm dis}\Theta^{-1}=\hat V_{\rm dis}$. These anti-unitary symmetries impose strong constraints on the field-free optical transition amplitudes. As a result, the symmetric and the antisymmetric rates have opposite symmetry properties: 
\begin{equation}
\big[W_{\alpha\beta}^{\updownarrow}\big]^{S,A}=\pm\big[W_{\bar\alpha\bar\beta}^{\updownarrow}\big]^{S,A},\quad\big[W_{\alpha\beta}^{\circpol}\big]^{S,A}=\mp\big[W_{\bar\alpha\bar\beta}^{\circpol}\big]^{S,A},
 \label{eq:symmetryinj}
\end{equation}
where $|\bar\alpha\rangle=\Theta|\alpha\rangle$ and $W_{\alpha\beta}^{\circpol}\equiv W_{\alpha\beta}^{\rm LH}-W_{\alpha\beta}^{\rm RH}$ is the circular transition rate \cite{AhnGuoNagaosa2020}. For linear polarization $\linpol$, the reciprocal rate is even under $\Theta$, whereas the anti-symmetric rate is odd. For the circular component, these parities are reversed because $\Theta$ interchanges the two optical helicities. 

In fact, this distinction between $W_{\alpha\beta}^{S,A}$ rates allows to distinguish the symmetry-allowed Hermitian (i.e. conventional photocurrents that survive when $\hat t^a=\hat r^a$) and non-Hermitian photocurrents (arises from $\mathcal Z_{\alpha\beta}^{ab}(\omega)$). A prime example are injection-type photocurrents~\cite{BelinicherSturman1980,AhnGuoNagaosa2020,SipeShkrebtii2000,WatanabeYanase2021}: bulk photocurrents formed from the velocity imbalance generated by optical interband transitions. By writing $W_{\alpha\to \beta}=W_{\alpha\beta}^S+W_{\alpha\beta}^A$, the injection current can be decomposed into Hermitian and non-Hermitian parts: $\mathbf j_{\rm inj}=\mathbf j_{\rm inj}^{\rm H}+\mathbf j_{\rm inj}^{\rm NH}$, with $\mathbf j_{\rm inj}^{\rm H}=\frac{e\tau}{2}\sum_{\alpha,\beta}f_{\alpha\beta}\mathbf v_{\beta\alpha}W_{\alpha\beta}^S$ and
\begin{equation}
 \mathbf j_{\rm inj}^{\rm NH}
 =\frac{e\tau}{2}\sum_{\alpha,\beta}\widetilde N_{\beta\alpha}
 f_{\alpha\beta}\mathbf v_{\beta\alpha}W_{\alpha\beta}^A,
 \,\, \widetilde N_{\beta\alpha}=2n(\epsilon_{\beta\alpha})+1,
 \label{eq:jinjSA}
\end{equation}
where $\mathbf v_{\beta\alpha}=\mathbf v_\beta-\mathbf v_\alpha$ is the velocity difference, $f_{\alpha\beta}=f_\alpha-f_\beta$ is the Fermi occupation difference, and $\epsilon_{\beta\alpha}=\epsilon_\beta-\epsilon_\alpha$ is the energy difference. Here $n(\epsilon)=[\exp(\epsilon/k_{\rm B}T)-1]^{-1}$ is the Bose function evaluated at the electronic temperature $T$, $e<0$ denotes the electron charge, and $\tau$ is the relaxation time. By applying Eq.~\eqref{eq:symmetryinj} to Eq.~\eqref{eq:jinjSA}, we find that $\mathbf j_{\rm inj}^{\rm H}$ and $\mathbf j_{\rm inj}^{\rm NH}$ have opposite anti-unitary symmetry transformation properties, as tabulated in Table~1. While $\mathbf j_{\rm inj}^{\rm H}$ produces the conventional injection current with the same symmetry properties as expected in the clean limit. 

Indeed, $[\mathbf{j}_{\rm inj}^{\rm NH}]^\updownarrow$ for linearly polarized light is the Ballistic current: an injection (velocity) type photocurrent that arises from the antisymmetric transition rate in Eq.~\eqref{eq:antirate} for linearly polarized light, see e.g., extensive discussion in Refs.~\cite{BelinicherSturman1988, BelinicherSturman1980, LeppenenGolub2023, Sturman2020, SturmanFridkin1992, DaiEtAl2021, Ivchenko, DaiRappe2021}. Unlike  $[\mathbf{j}_{\rm inj}^{\rm H}]^\updownarrow$ that vanishes for $\mathcal{T}$ symmetry (i.e. non-magnetic systems), $[\mathbf{j}_{\rm inj}^{\rm NH}]^\updownarrow$ is non-vanishing under $\mathcal{T}$ symmetry and can provide substantial contributions to the photocurrent in non-magnetic systems \cite{DaiEtAl2021, DaiRappe2021, BelinicherSturman1988, BelinicherSturman1980, LeppenenGolub2023, Sturman2020, SturmanFridkin1992, Ivchenko} Nevertheless, such ballistic currents vanish in $\mathcal{PT}$ symmetric systems (see Table~1). 

Instead, we find that circular non-Hermitian injection currents $[\mathbf{j}_{\rm inj}^{\rm NH}]^{\circpol}$ persist in $\mathcal{PT}$ symmetric magnets (Table~1); these arise from antisymmetric optical transition rates for circularly polarized electromagnetic fields.  
The circular contribution in Eq.~\eqref{eq:antirate} is selected by the circular field factor $H^{ab}(\omega)$. This contrasts with the conventional circular injection current $[\mathbf{j}_{\rm inj}^{\rm H}]^{\circpol}$ that vanishes in $\mathcal{PT}$ symmetric systems \cite{AhnGuoNagaosa2020, WatanabeYanase2021}. Indeed, this vanishing of $[\mathbf{j}_{\rm inj}^{\rm H}]^{\circpol}$ in $\mathcal{PT}$ symmetric systems is sometimes used to delineate circular shift photocurrent contributions from those that are injection type \cite{KangEtAl2025}. $[\mathbf{j}_{\rm inj}^{\rm NH}]^{\circpol}$, on the other hand, demonstrates that injection type processes can also produce circular photocurrents in $\mathcal{PT}$ symmetric materials. 

\section{Non-Hermitian Shift Photocurrents}
\begin{table*}[t]
\centering
\label{tab:photocurrent}
\renewcommand{\arraystretch}{1.35}
\begin{tabular*}{\textwidth}
  {@{\extracolsep{\fill}}llcccc@{}}
\toprule
\textbf{Category} & \textbf{Mechanism}
  & $\boldsymbol{\mathcal{T}}$ & $\boldsymbol{\mathcal{PT}}$ & \textbf{Geometric Origin} & \textbf{Reference}\\
\midrule
Hermitian
  & Injection current (\linpol/\circpol)
  & \nomark(\yesmark) & \yesmark(\nomark) & Symmetric quantum metric &{\cite{AhnGuoNagaosa2020, WatanabeYanase2021, SipeShkrebtii2000, DeJuan, Parker}} \\ photocurrent
  & Shift current (\linpol/\circpol)
  & \yesmark(\nomark) & \nomark(\yesmark) & Symmetric connection &{\cite{AhnGuoNagaosa2020, WatanabeYanase2021, SipeShkrebtii2000, vonBaltzKraut1981, AhnEtAl2022, Parker}}\\
\addlinespace[0.7em]
Non-Hermitian
  & Ballistic current
  & \yesmark & \nomark  & Skew quantum metric &{\cite{BelinicherSturman1988, BelinicherSturman1980, LeppenenGolub2023, Sturman2020, SturmanFridkin1992, Ivchenko}}\\ photocurrent 
  & Circular non-Hermitian injection current
  & \nomark & \yesmark & Skew quantum metric &{Eq.~\eqref{eq:jinjSA}}\\
  & Linear non-Hermitian shift current
  & \nomark & \yesmark & Skew quantum connection &{Eq.~\eqref{eq:shiftcurrent2}}\\
  & Circular non-Hermitian shift current
  & \yesmark & \nomark & Skew quantum connection &{Eq.~\eqref{eq:shiftcurrent2}} \\
\bottomrule
\end{tabular*}
\caption{Hermitian and Non-Hermitian Photocurrent mechanisms and constraints imposed by $\mathcal T$ and $\mathcal{PT}$ symmetries. \linpol\ and \circpol\ denote linear and circular polarization, respectively. Hermitian photocurrents reduce to conventional clean photocurrents in the clean limit $\hat{t} \to \hat{r}$. For these two rows, the \nomark\ and \yesmark\ symbols outside and inside parentheses correspond to linear \linpol\ and circular \circpol\ polarization, respectively. Non-Hermitian photocurrents are described in the last four rows and have opposite symmetry properties to their clean counterparts. Here the Ballistic photocurrent refers to injection type photocurrents produced by the antisymmetric optical transition rate under linear polarization light.}
\end{table*}

We now explain how non-Hermitian optical transitions and the non-Hermitian interference Wilson-loop tensor $\mathcal Z_{\alpha\beta}^{ab}$ dramatically change  geometric phase phenomena. To do so, we turn to examine the shift current that arises from a real-space coordinate shift (and electric polarization) upon optical absorption; a constant pumping of this transition produces a current \cite{vonBaltzKraut1981,MorimotoNagaosa2016,SipeShkrebtii2000,AhnGuoNagaosa2020,WatanabeYanase2021}. In a $T$-matrix formulation, the shift current can be readily computed by examining the real-space shift vector that electronic wavepackets accrue as they scatter from $\alpha\to\beta$ \cite{SinitsynNiuMacDonald2006,Xiong2022Thesis,ShiSong2019} (see detailed derivation for interband optical transitions in {\bf SM}):
\begin{equation}
 R_{\alpha\to\beta}^{\mu,(\omega)}
 =\mathcal A_\beta^\mu-\mathcal A_\alpha^\mu
 -[\partial_\beta^\mu+\partial_\alpha^\mu]
 \operatorname{arg}T_{\beta\alpha}^{(\omega)},
 \label{eq:shift}
\end{equation}
where $\mathcal A_\alpha=i\langle u_\alpha|\nabla_\alpha u_\alpha\rangle$ is the Berry connection of state $\alpha$, and $\partial_\alpha^\mu$ denotes the partial derivative with respect to the $\mu$ component of the momentum associated with state $\alpha$, holding the momentum of state $\beta$ fixed. The derivative $\partial_\beta^\mu$ is defined analogously.

In the clean limit $\hat t^a=\hat r^a$, the shift vector reduces to the familiar expression $R_{n\to m}^{\mu,(\omega)}(\mathbf k)=\mathcal A_m^\mu-\mathcal A_n^\mu-\partial_{k^\mu}\operatorname{arg}[V_\omega(\mathbf k)]_{mn}$ \cite{vonBaltzKraut1981,SipeShkrebtii2000,MorimotoNagaosa2016}. Since $\hat r^a$ is Hermitian and the reverse process $\beta \to \alpha$ uses the conjugate electromagnetic field, the clean amplitudes satisfy $T_{\alpha\beta}^{(0)}=(T_{\beta\alpha}^{(0)})^*$, giving $R_{\alpha\to\beta}^{(0)}=-R_{\beta\to\alpha}^{(0)}$. This means that $R_{\alpha\to\beta}^{(0)}$ has an antisymmetric character.

However, when $\hat t^a\ne(\hat t^a)^\dagger$, the phases accrued by scattering can render $R_{\alpha\to\beta}^{(\omega)}+R_{\beta\to\alpha}^{(-\omega)}$ non-vanishing. To see this, we decompose the generalized $T$-matrix shift vector into
\begin{equation}
R_{\alpha\to\beta}^{(\omega)}\hspace{-0.5mm}=\hspace{-0.5mm}R_{\alpha\beta}^{S,(\omega)}\hspace{-0.5mm}+\hspace{-0.5mm}R_{\alpha\beta}^{A,(\omega)}\hspace{-2mm},
 \,\,\,
 R_{\alpha\beta}^{S,A,(\omega)}
 =\frac{R_{\alpha\to\beta}^{(\omega)}\pm R_{\beta\to\alpha}^{(-\omega)}}{2},
 \label{eq:shift-SA-decomposition}
\end{equation}
where $S,A$ describe symmetric and antisymmetric shift vectors under exchange $\alpha\leftrightarrow\beta$. Notice that $R_{\alpha\beta}^A$ has the same exchange properties as $R_{\alpha\to\beta}^{(0)}$ and reduces to it in the clean limit. In contrast, using Eq.~\eqref{eq:shift}, we find
\begin{equation}
 [R_{\alpha\beta}^\mu]^{S,(\omega)}
 =-\frac{1}{2}[\partial_\beta^\mu+\partial_\alpha^\mu]
 \operatorname{arg}(T_{\beta\alpha}^{(\omega)}T_{\alpha\beta}^{(-\omega)}).
 \label{Rsymmetric}
\end{equation}
In the clean limit, $T_{\beta\alpha}^{(0)}T_{\alpha\beta}^{(0)}=|T_{\beta\alpha}^{(0)}|^2$, hence $R_{\alpha\beta}^S$ vanishes. However, phases associated with multiple scattering in the absorption process turn on $R_{\alpha\beta}^{S,(\omega)}$ to enrich the behavior of the shift current. Importantly, this symmetric shift vector is directly related to $\mathcal Z_{\alpha\beta}^{ab}(\omega)$. Decomposing into linear and circular contributions, we obtain: 
\begin{align}
 T_{\beta\alpha}^{(\omega)}&T_{\alpha\beta}^{(-\omega)}\hspace{-1mm}=\Big\{
 L^{ab}_\omega\Big[
 \operatorname{Re}\!\left(
 \mathcal X_{\alpha\beta}^b\mathcal X_{\beta\alpha}^a
 \hspace{-1mm}-\hspace{-1mm}\mathcal Y_{\alpha\beta}^b\mathcal Y_{\beta\alpha}^a
 \right)^\omega
 \hspace{-1mm}+\hspace{-1mm}2i\operatorname{Re}\mathcal Z_{\alpha\beta}^{ab,\omega}
 \Big] \nonumber \\
 &\quad-H^{ab}_\omega\Big[
 \operatorname{Im}\!\left(
 \mathcal X_{\alpha\beta}^b\mathcal X_{\beta\alpha}^a
 \hspace{-1mm}-\hspace{-1mm}\mathcal Y_{\alpha\beta}^b\mathcal Y_{\beta\alpha}^a
 \right)^\omega
 \hspace{-1mm}+\hspace{-1mm}2i\operatorname{Im}\mathcal Z_{\alpha\beta}^{ab,\omega}
 \Big]\Big\}e^2,
\label{Non-Hermitian Detachment}
\end{align}
where $(\mathcal X_{\alpha\beta}^b\mathcal X_{\beta\alpha}^a-\mathcal Y_{\alpha\beta}^b\mathcal Y_{\beta\alpha}^a)^\omega \equiv \mathcal X_{\alpha\beta}^{b,-\omega}\mathcal X_{\beta\alpha}^{a,\omega}-\mathcal Y_{\alpha\beta}^{b,-\omega}\mathcal Y_{\beta\alpha}^{a,\omega}$. In fact, Eq.~(\ref{Non-Hermitian Detachment}) demonstrates that finite $\mathcal{Z}^{ab}_{\alpha\beta}(\omega)$ is critical to produce $[R_{\alpha\beta}^\mu]^{S,(\omega)}$. Take linearly polarized light for instance: if $\operatorname{Re}\mathcal Z_{\alpha\beta}^{ab}(\omega) = 0$, we find $ \operatorname{arg}(T_{\beta\alpha}^{(\omega)}T_{\alpha\beta}^{(-\omega)})=0$ for all $\mathbf{k}$. Similarly, for circularly polarized light, the circular contribution to the imaginary part of $T_{\beta\alpha}^{(\omega)}T_{\alpha\beta}^{(-\omega)}$ vanishes when $\operatorname{Im}\mathcal Z_{\alpha\beta}^{ab}(\omega)=0$. Consequently, the symmetric shift vector $[R_{\alpha\beta}^\mu]^S$ induced by either linearly or circularly polarized light originates from the complex-valued non-Hermitian interference Wilson-loop tensor $\mathcal Z_{\alpha\beta}^{ab}$.

With the above knowledge, we can illustrate the impact of the non-Hermitian interference Wilson-loop tensor $\mathcal Z_{\alpha\beta}^{ab}$ on the shift current by computing the rate of change of electric polarization as $\mathbf{j}_{\rm shift} = \mathbf{j}_{\rm shift}^{\rm H} +  \mathbf{j}_{\rm shift}^{\rm NH}$:
\begin{align}
    \mathbf{j}_{\rm shift}^{\rm H} & \hspace{-1mm}= \frac{e}{2}\hspace{-2mm}\sum_{\alpha,\beta,\pm\omega} \hspace{-2mm}f_{\alpha \beta} ( W^{S,(\omega)}_{\alpha \beta} R^{A,(\omega)}_{\alpha \beta} \hspace{-1mm}+ R^{S,(\omega)}_{\alpha \beta} W^{A,(\omega)}_{\alpha \beta}), 
\label{eq:shiftcurrent1}\\
    \mathbf{j}_{\rm shift}^{\rm NH} & \hspace{-1mm}= \frac{e}{2}\hspace{-2mm}\sum_{\alpha,\beta,\pm\omega} \hspace{-2mm}\widetilde{N}_{\beta\alpha} f_{\alpha \beta} (R^{S,(\omega)}_{\alpha \beta} W^{S,(\omega)}_{\alpha \beta} \hspace{-1mm}+ W^{A,(\omega)}_{\alpha \beta} R^{A,(\omega)}_{\alpha \beta}). 
\label{eq:shiftcurrent2}
\end{align}
Eq.~\eqref{eq:shiftcurrent1} and ~\eqref{eq:shiftcurrent2} are one of the main results of our work. They demonstrate how non-Hermitian processes directly affect the shift current -- a geometric phase phenomenon -- producing a non-Hermitian shift current. 

In the clean limit, only the $W^S_{\alpha\beta}R^A_{\alpha\beta}$
term in Eq.~\eqref{eq:shiftcurrent1} survives, which yields the conventional shift current \cite{vonBaltzKraut1981, SipeShkrebtii2000, MorimotoNagaosa2016}. The remaining terms arise from multiple scattering processes in the optical absorption event which describe impurity-induced corrections. Meanwhile, the different occupation factors between Hermitian and non-Hermitian follow from combining forward and reverse transitions. It is interesting to note that both the non-Hermitian shift current in Eq.~\eqref{eq:shiftcurrent2} as well as the non-Hermitian injection current in Eq.~\eqref{eq:jinjSA} possess a factor of $\widetilde{N}(\epsilon_{\beta\alpha})$. This term arises because the summation of the forward and reverse Pauli occupation factors satisfies $f_\alpha(1-f_\beta)+f_\beta(1-f_\alpha)=\widetilde N(\epsilon_{\beta\alpha})f_{\alpha\beta}$.

Notice that $\mathbf j_{\rm shift}^{\rm H}$ contains both the conventional clean-limit term and the impurity correction $R^S_{\alpha\beta}W^A_{\alpha\beta}$, while
$\mathbf j_{\rm shift}^{\rm NH}$ represents a completely new contribution enabled by the non-Hermiticity of the transition operator. As we show below, it has opposite anti-unitary symmetry parities to $\mathbf j_{\rm shift}^{\rm H}$ for each polarization of the driving electromagnetic field.

To see this, notice that $R^{S,A}$ have different symmetry properties. Under $\mathcal T$ transformation, the symmetry-related states $|\bar\alpha\rangle=\Theta_{\mathcal T}|\alpha\rangle$ have reversed momenta. In particular, $\boldsymbol{\mathcal A}_{\bar\alpha}=\boldsymbol{\mathcal A}_\alpha$ and $\boldsymbol{\partial}_\alpha^\mu = -\boldsymbol{\partial}_{\bar{\alpha}}^\mu$ 
with analogous relations for $\beta$ (see {\bf SM}). For the field-free amplitudes, time-reversal symmetry gives $t^a_{\bar\alpha\bar\beta}(\omega)=t^a_{\beta\alpha}(-\omega)$, while the anti-unitary transformation $L^{ab}+iH^{ab}\to L^{ab}-iH^{ab}$ interchanges \textbf{LCP} and \textbf{RCP} light. Contracting these amplitudes with the electromagnetic field produces the relations: 
$T^{\updownarrow}_{\bar\alpha\bar\beta}=T^{\updownarrow}_{\beta\alpha}$, $T^{\rm LH}_{\bar\alpha\bar\beta}=T^{\rm RH}_{\beta\alpha}$, and $T^{\rm RH}_{\bar\alpha\bar\beta}=T^{\rm LH}_{\beta\alpha}$. Combining these relations gives the $\mathcal T$-symmetry constraints for the shift vector
\begin{equation}
 \big[R_{\alpha\beta}^\updownarrow\big]^{S,A}
 =\mp\big[R_{\bar\alpha\bar\beta}^\updownarrow\big]^{S,A},
 \, \, 
 \big[R_{\alpha\beta}^{\rm LH}\big]^{S,A}
 =\mp\big[R_{\bar\alpha\bar\beta}^{\rm RH}\big]^{S,A}, 
 \label{eq.RTsymmetry}
\end{equation}
where for linear polarization, the symmetric shift vector is $\mathcal T$ odd, while the antisymmetric shift vector is $\mathcal T$ even. For circular polarization, the anti-unitary transformation exchanges the helicity-resolved shift vectors $R^{\rm LH}$ and $R^{\rm RH}$, reversing these parities for the helicity-odd combination.

The same analysis can be performed for $\mathcal{PT}$ symmetric systems. Under $\mathcal{PT}$ anti-unitary transformation, the partner states $|\tilde\alpha\rangle=\Theta_{\mathcal{PT}}|\alpha\rangle$ retain their momenta. The corresponding relations are $\boldsymbol{\mathcal A}_{\tilde\alpha}=-\boldsymbol{\mathcal A}_\alpha$ and $\boldsymbol{\partial}_\alpha^\mu = \boldsymbol{\partial}_{\tilde{\alpha}}^\mu$ (see {\bf SM}). Since the dipole operator is $\mathcal{PT}$ odd, the field-free amplitudes satisfy $t^a_{\tilde\alpha\tilde\beta}(-\omega)=-t^a_{\beta\alpha}(\omega)$. This overall minus sign contributes only a momentum-independent phase and therefore does not affect the shift vector. Using the same reasoning as above we find: 
$T^{\updownarrow}_{\tilde\alpha\tilde\beta}=-T^{\updownarrow}_{\beta\alpha}$,$T^{\rm LH}_{\tilde\alpha\tilde\beta}=-T^{\rm RH}_{\beta\alpha}$, and $T^{\rm RH}_{\tilde\alpha\tilde\beta}
=-T^{\rm LH}_{\beta\alpha}$, which yield $\mathcal{PT}$ constraints 
\begin{equation}
 \big[R_{\alpha\beta}^\updownarrow\big]^{S,A}
=\pm\big[R_{\tilde\alpha\tilde\beta}^\updownarrow\big]^{S,A},\,\,
 \big[R_{\alpha\beta}^{\rm LH}\big]^{S,A}
 =\pm\big[R_{\tilde\alpha\tilde\beta}^{\rm RH}\big]^{S,A}.
 \label{eq.RPTsymmetry}
\end{equation}
We can see that, for linear polarization, the symmetric shift vector is $\mathcal{PT}$ even and the antisymmetric shift vector is $\mathcal{PT}$ odd. For circular polarization, the anti-unitary transformation again exchanges the helicity-resolved shift vectors $R^{\rm LH}$ and $R^{\rm RH}$, reversing these parities for the helicity-odd combination. Combined with the transition-rate symmetries in Eq.~\eqref{eq:symmetryinj}, these relations determine the allowed polarization process of the Hermitian and non-Hermitian shift currents under anti-unitary symmetries.

Inserting Eqs.~\eqref{eq.RTsymmetry},\eqref{eq.RPTsymmetry} and Eq.~\eqref{eq:symmetryinj} into Eq.~\eqref{eq:shiftcurrent1},~\eqref{eq:shiftcurrent2} we find the symmetry properties in Table I (rows). Notice that $\mathbf{j}_{\rm shift}^{\rm H}$ has the same symmetry properties as the clean shift current \cite{AhnGuoNagaosa2020}: namely, the linearly polarized $\mathbf{j}_{\rm shift}^{\rm H}$ is time reversal even but $\mathcal{PT}$ odd and the circularly polarized $\mathbf{j}_{\rm shift}^{\rm H}$ is time reversal odd but $\mathcal{PT}$ even. In contrast, $\mathbf{j}_{\rm shift}^{\rm NH}$ is different: the linearly polarized $\mathbf{j}_{\rm shift}^{\rm NH}$ is time reversal odd but $\mathcal{PT}$ even and the circularly polarized $\mathbf{j}_{\rm shift}^{\rm NH}$ is time reversal even but $\mathcal{PT}$ odd. These depend on the non-Hermitian processes we identified above and appear only beyond the clean limit when multiple scattering processes participate in the absorption event. 

As we will see, the circular non-Hermitian shift current is particularly striking in non-magnetic excitonic systems. This is because unlike non-interacting free electrons, excitonic systems possess zero injection currents~\cite{YangSrivastavaSong2026}. This makes circular skew scattering shift current the leading order circular photocurrent for excitons in time reversal symmetric systems. A detailed numerical example and full discussion will follow in Sec.~$\mathbf{V}$. 

\section{Non-Hermitian Quantum Geometry of Optical Transitions}

The non-Hermitian optical transitions and photocurrents depart sharply from the conventional electric-dipole description of optical processes \cite{SipeShkrebtii2000, AhnGuoNagaosa2020, vonBaltzKraut1981, MorimotoNagaosa2016, WatanabeYanase2021, AhnEtAl2022, Parker, DeJuan}. As we now explain, these processes are naturally captured within a non-Hermitian quantum geometry of optical transitions. To see this, we first examine the transition vectors describing non-Hermitian optical transitions from $|u_n(\mathbf k)\rangle$ to $|u_m(\mathbf k')\rangle$:
\begin{equation}
 \hat e_{mn}^a(\mathbf k',\mathbf k)
 =t_{mn}^a(\mathbf k',\mathbf k)
 |u_m(\mathbf k')\rangle\langle u_n(\mathbf k)|.
 \label{eq:transition-section'}
\end{equation}
Here $t_{mn}^a(\mathbf k',\mathbf k)=\langle u_m(\mathbf k')|\hat t^a|u_n(\mathbf k)\rangle$ is the field-free transition matrix element, and $\mathbf k$ and $\mathbf k'$ are the initial and final momenta respectively. Here we use $\alpha\equiv(n,\mathbf k)$ and $\beta\equiv(m,\mathbf k')$, which explicitly manifests the continuous variables $\mathbf k,\mathbf k'$ that construct the product base manifold $\mathcal M=K_i\times K_f$ of the non-Hermitian quantum fiber bundle that we focus on. Notice that Eq.~\eqref{eq:transition-section'} directly mirrors the tangent vectors described in Ref.~\cite{AhnEtAl2022} when $\hat{t}^a = \hat{r}^a$ and $\mathbf{k} = \mathbf{k}'$. In that case, this describes {\it direct} resonant optical transitions~\cite{AhnEtAl2022}. Importantly in our case we focus on {\it indirect} transitions where $\mathbf{k}$ need not be the same as $\mathbf{k}'$ as well as using the optical transitions described by a $T$-matrix that in general is non-Hermitian, i.e. $\hat{t}^a \neq (\hat{t}^\dagger)^a$. In what follows, we will drop the explicit $\mathbf{k}, \mathbf{k'}$ dependence since $n$ is always attached to $\mathbf{k}$ and $m$ is always attached to $\mathbf{k'}$.

Indeed, this distinction allows us to separate the field free $T$-matrix into Hermitian and non-Hermitian  parts $\hat{t}^a=\hat {\mathcal X^a}+i\hat {\mathcal Y^a}$. This property juxtaposes with that of the simple geometry found in Ref.~\cite{AhnEtAl2022} where electric dipole light-matter interaction ($\hat{\mathbf{r}} = \hat{\mathbf{r}}^\dagger$) that ensures $r_{mn}^a = (r_{nm}^a)^*$. 

This difference in $\hat{t}^a$ and $\hat{\mathbf{r}}^a$ have important consequences for their quantum geometry and observables. For example, the electric-dipole-induced interband quantum metric $[Q^{(0)}]_{mn}^{ab}=r_{nm}^b r_{mn}^a=([Q^{(0)}]_{nm}^{ab})^*$ gives equal transition rates for forward ($n\to m$) and reverse ($m\to n$) processes. 

In contrast, for non-Hermitian transitions, $t_{mn}^a$ need not equal $(t_{nm}^a)^*$. The corresponding $T$-matrix-induced optical transition quantum metric is
\begin{equation}
 Q_{mn}^{ab}
 =(\hat e_{mn}^b,\hat e_{mn}^a)
 =(t_{mn}^b)^*t_{mn}^a
 =[t^\dagger]^b_{nm}t_{mn}^a,
 \label{eq:transition-quantum-metric}
\end{equation}
where $(A,B)=\operatorname{Tr}(A^\dagger B)$ is the Hilbert--Schmidt inner product \cite{AhnEtAl2022}, and $[t^\dagger]^b_{nm}=\langle u_n(\mathbf k)|(\hat t^b)^\dagger|u_m(\mathbf k')\rangle=(t_{mn}^b)^*$. We have otherwise suppressed the momentum and frequency arguments. Non-Hermitian transition operators can therefore yield distinct quantum metrics for forward and reverse processes, $Q_{mn}^{ab}\ne(Q_{nm}^{ab})^*$, sharply contrasting with that of the conventional geometry for electric-dipole optical transitions. This distinction is one signature of the non-Hermitian quantum geometry of optical transitions.

On a physical level, because the quantum metric tracks the optical transition rates \cite{AhnEtAl2022}, this means that forward and reverse rates are different mirroring the antisymmetric rates discussed in Section~$\mathbf I$; notice that $Q_{mn}^{ab} \to [Q^{(0)}]_{mn}^{ab}$ when $\hat{t}^a \to \hat{\mathbf{r}}^a$. Importantly, $Q_{mn}^{ab}$ can be naturally separated into two parts:
\begin{equation}
 Q_{mn}^{ab}
 =[Q_{\rm S}]_{mn}^{ab}+[Q_{\rm A}]_{mn}^{ab},
 \quad
 [Q_{\rm S,A}]_{mn}^{ab}
 =\frac{Q_{mn}^{ab}\pm(Q_{nm}^{ab})^*}{2}.
 \label{eq.QMSplit}
\end{equation}
Here $[Q_{\rm S}]_{mn}^{ab}=\mathcal X_{nm}^b\mathcal X_{mn}^a+\mathcal Y_{nm}^b\mathcal Y_{mn}^a$ is the symmetric metric, while $[Q_{\rm A}]_{mn}^{ab}=i[\mathcal Z_{nm}^{ab}-(\mathcal Z_{nm}^{ba})^*]$ is the skew quantum metric. Notice that the symmetric quantum metric is generated by $\mathcal X_{nm}^b\mathcal X_{mn}^a+\mathcal Y_{nm}^b\mathcal Y_{mn}^a$ and the skew quantum metric is generated by the non-Hermitian interference loop tensor $\mathcal Z_{mn}^{ab}$. 

Unlike the conventional Hermitian quantum geometry \cite{AhnEtAl2022, Varolin2019HilbertSpaceTechniques}, $[Q_{\rm A}]_{mn}^{ab}$ is a non-Hermitian quantum metric (pseudo Hermitian metric \cite{Bognar1974IndefiniteInnerProduct}) that directly captures the non-Hermicity of transition processes since it vanishes when $\hat{t}^a \to \hat{\mathbf{r}}^a$ becomes Hermitian. Indeed, this skew quantum metric tracks the antisymmetric interband transition rate in Eq.~\eqref{eq:antirate} and produces the ballistic and circular non-Hermitian injection currents in Eq.~\eqref{eq:jinjSA}. 

As a result, using Eq.~\eqref{eq.QMSplit} we can directly write down the second-order injection photocurrents as 
$j_{\mathrm{inj}}^{c}=\sum_{\pm\omega,m,n}[\sigma_{\mathrm{inj}}(\omega)]_{mn}^{c;ab}E^a(\omega)E^b(-\omega)$, where $\sigma_{\mathrm{inj}} = \sigma_{\mathrm{inj}}^{\mathrm{H}}+ \sigma_{\mathrm{inj}}^{\mathrm{NH}}$  
is the second-order optoelectric conductivity given by
\begin{align}
[\sigma_{\mathrm{inj}}^{\mathrm{H}}]_{mn}^{c;ab}
 &=\frac{\pi e^3\tau}{\hbar V}
 \sum_{k,k'}\Delta_\omega f_{nm}\mathbf v^c_{mn}
 [Q_{\rm S}]_{mn}^{ab},
 \label{eq:normal-injection}
 \\
 [\sigma_{\mathrm{inj}}^{\mathrm{NH}}]_{mn}^{c;ab}
 &=\frac{\pi e^3\tau}{\hbar V}
 \sum_{k,k'}\Delta_\omega\widetilde N_{mn}
 f_{nm}\mathbf v^c_{mn}
 [Q_{\rm A}]_{mn}^{ab}.
 \label{eq:normal-skew-injection}
\end{align}
Here the symmetric and skew quantum metrics encode the quantum geometric origins of the Hermitian and non-Hermitian injection response tensors, respectively. In particular, the non-Hermitian response tensor originates from the non-Hermitian interference loop tensor $\mathcal Z_{mn}^{ab}$ as well as the non-Hermitian property of this quantum geometry.

\begin{figure}
\centering
\includegraphics[width=1\linewidth]{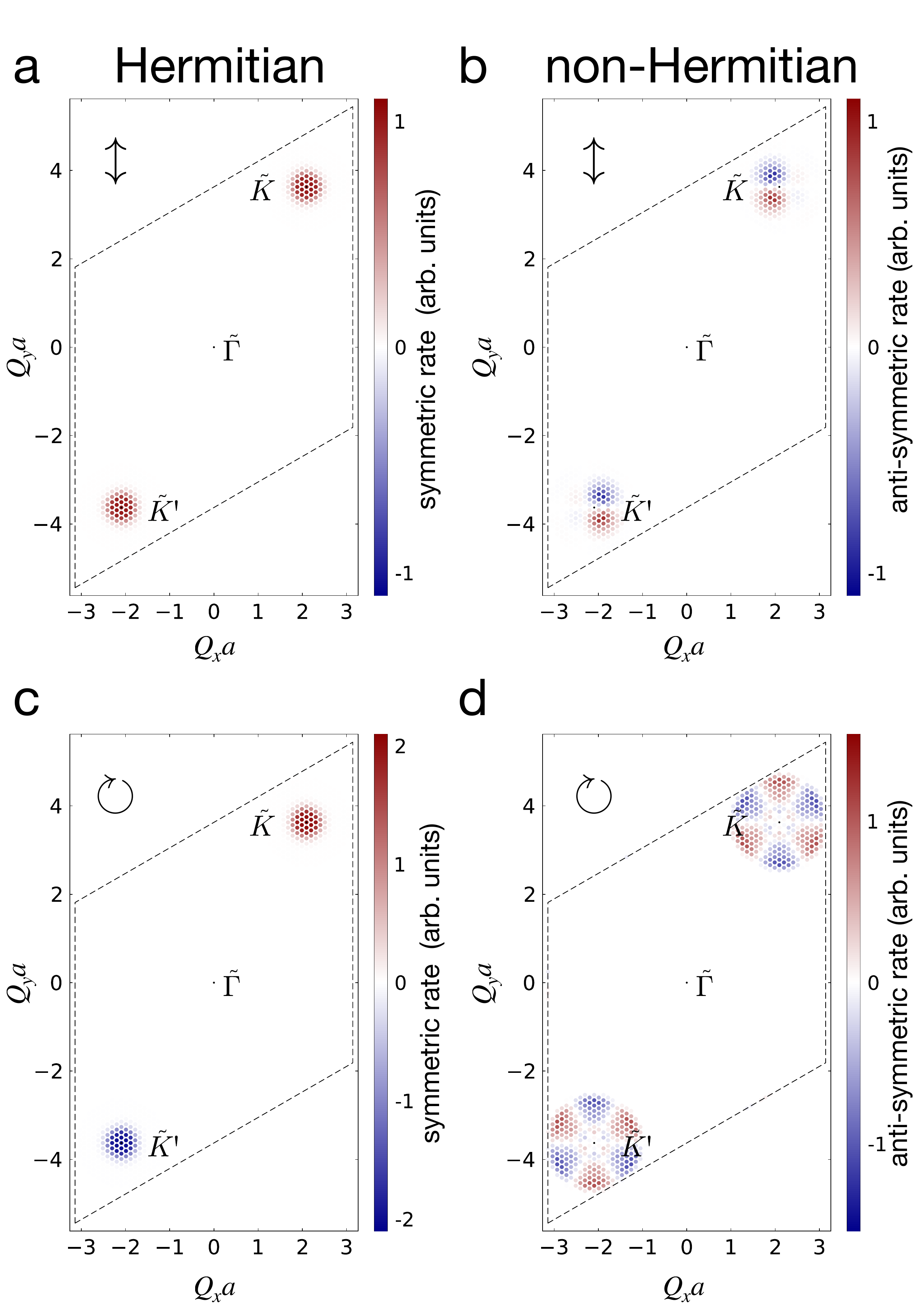}
\caption{{\bf Symmetric and anti-symmetric rates for linearly and circularly polarized light.} The disorder-assisted transition rate $W_{0\rightarrow n_{ex,\bm{Q}}}$ from ground state $0$ to an exciton state $n_{ex}$ with center-of-mass $\bm{Q}$ is calculated using a 3-band model of MoS$_2$. Here $\bm{Q}$ is the momentum characterizing the indirect optical transition. See Sec.{\bf V} for calculation details. The results in all panels are obtained on a $60\times 60$ lattice. The first column illustrates the symmetric rates governed by Hermiticity, and the second column shows the anti-symmetric rates arising from non-Hermiticity. {\bf a-b.} Momentum-resolved linear transition rates. The linear symmetric rate in $\bf{a}$ is even under time-reversal $\mathcal{T}$ that maps $\bm{Q}\rightarrow -\bm{Q}$ , whereas the linear anti-symmetric rate in $\bf{b}$ is odd under $\mathcal{T}$ . {\bf c-d.} Momentum-resolved circular transition rates. These transformation properties under $\mathcal{T}$ are reversed: the circular symmetric rate is odd under $\mathcal{T}$ in $\bf{c}$, whereas the circular anti-symmetric rate is even in $\bf{d}$. These results are consistent with the symmetry properties shown in Eq.~\eqref{eq:symmetryinj}.}
\label{fig:rate}
\end{figure}

Indeed, the distinction between symmetric and skew quantum geometries directly manifest in the corresponding transition rates. As an example, we plot the momentum-resolved linear and circular transition rates in Fig.~\ref{fig:rate}. For the linearly polarized transition rates, the symmetric contribution is even under $\mathcal{T}$ which flips the momentum of the indirect optical transition: $\bm{Q}\rightarrow -\bm{Q}$ (Fig.~\ref{fig:rate}(a)), whereas the antisymmetric contribution is odd under $\mathcal{T}$ (Fig.~\ref{fig:rate}(b)). These properties are reversed for circular transition rates: the circular symmetric transition rate is odd under $\mathcal{T}$ (Fig.~\ref{fig:rate}(c)), whereas the circular antisymmetric transition rate is even under $\mathcal{T}$ (Fig.~\ref{fig:rate}(d)). This aligns with the symmetry relations in Eq.~\eqref{eq:symmetryinj}. 

We now turn to the quantum connection $C_{mn}^{acb}=(\hat e_{mn}^b,\nabla^c\hat e_{mn}^a)$ \cite{AhnEtAl2022}, written in terms of the field-free transition matrix elements as
\begin{equation}
 C_{mn}^{acb}
 =(t_{mn}^b)^*\mathcal D_c^{mn}t_{mn}^a.
 \label{eq:transition-connection-fieldfree}
\end{equation}
Here $\nabla^c$ acts on the transition sections, while $\mathcal D_c^{mn}=\partial_{k^c}+\partial_{k'^c}-i\bigl[\mathcal A_m^c(\mathbf k')-\mathcal A_n^c(\mathbf k)\bigr]$ is the corresponding covariant derivative acting on their coefficients.

In the same fashion as Eq.~\eqref{eq.QMSplit}, the quantum connection for the non-Hermitian optical transitions can be similarly split into 
\begin{equation}
    C_{mn}^{acb} = [C_{\rm S}]_{mn}^{acb} + [C_{\rm A}]_{mn}^{acb}, \quad [C_{\rm S,A}]_{mn}^{acb}  = \frac{C_{mn}^{acb} \pm [C_{nm}^{acb}]^*}{2},
\label{SAconnection}
\end{equation}
where $[C_{\rm S}]_{mn}^{acb}$ and $[C_{\rm A}]_{mn}^{acb}$ are the symmetric and skew quantum connections \cite{Bognar1974IndefiniteInnerProduct} of the optical transition, compatible with the corresponding symmetric and skew quantum metrics, respectively. Similar to that described in Eq.~\eqref{Rsymmetric}, $[C_{\rm A}]_{mn}^{acb}$ directly captures non-Hermitian processes since it vanishes when $\hat{t}^a \to \hat{\mathbf{r}}^a$ becomes Hermitian.

This allows us to write the Hermitian and non-Hermitian shift conductivities directly in terms of the symmetric and skew connections:
\begin{align}
 [\sigma_{\mathrm{shift}}^{\mathrm{H}}]_{mn}^{c;ab}
 &=\frac{i\pi e^3}{2\hbar V}
 \sum_{k,k'}\Delta_\omega f_{nm}
 \bigl([C_{\rm S}]_{mn}^{acb} - ([C_{\rm S}]_{mn}^{bca})^*
 \bigr),
 \label{eq:symmetric-shift-conductivity}\\
 [\sigma_{\mathrm{shift}}^{\mathrm{NH}}]_{mn}^{c;ab}
 &=\frac{i\pi e^3}{2\hbar V}
 \sum_{k,k'}\Delta_\omega
 \widetilde N_{mn}f_{nm}
 \bigl([C_{\rm A}]_{mn}^{acb} - ([C_{\rm A}]_{mn}^{bca})^*
 \bigr),
 \label{eq:skew-shift-conductivity}
\end{align}
The symmetric and skew quantum connections encode the geometric origins of the Hermitian and non-Hermitian shift response tensors, respectively. In particular, the non-Hermitian response arises from the non-Hermitian response tensor originates from the non-Hermitian interference loop tensor $\mathcal Z_{mn}^{ab}$ and its covariant momentum derivatives. It therefore probes the momentum dependence of non-Hermitian transition geometry; this mirrors how the skew quantum metric governs the non-Hermitian injection response. As a result, we directly attribute the non-Hermitian photocurrents to the skew quantum geometry described in Eq.~\eqref{eq.QMSplit} and Eq.~\eqref{SAconnection}.

It is interesting to note the clear separation of the symmetric and skew quantum metric and their compatible quantum connections in Eq.~\eqref{eq.QMSplit} and \eqref{SAconnection}. Skew metrics and skew connections vanish in the Hermitian limit and are formed in the same way for both metric and connection. On the other hand, taken together, Eqs.~\eqref{eq:antirate}, \eqref{Rsymmetric}, and \eqref{eq:shiftcurrent2} show that the Wilson-loop interference tensor $\mathcal Z_{mn}^{ab}$ enters both the antisymmetric transition rate and the symmetric shift vector contributing to the non-Hermitian shift current, revealing their common geometric origin in the asymmetry between forward and reverse optical transitions. At a deeper level, the distinction between skew and symmetric quantum geometry can be understood by comparing the quantum geometry of the optical transition $\hat e_{mn}^a$ basis vectors with that of its dual vectors $[\hat e^\dagger]^a_{mn}$. When $\hat{t}^a \to \hat{\mathbf{r}}^a$ for Hermitian optical transitions, the quantum geometry of the optical transition basis vectors and its dual space spanned by $[\hat e^\dagger]^a_{mn}$ are identical (they have identical geometric structures quantum metric and connection). For non-Hermitian processes captured by $\hat{t}^a$, the geometries in these two spaces diverge from each other producing the skew quantum geometry we describe above. This delineation underscores the rich structure of optical transition quantum geometry beyond the familiar electric-dipole processes to include non-Hermitian optical transition processes. 

\section{Disorder-induced Non-Hermitian Photocurrents}

To illustrate the non-Hermitian photocurrents and non-Hermitian optical transitions, we evaluate the
transition rates and photocurrents for scalar short-range disorder
and electric-dipole coupling to monochromatic light $\hat V(t)=\hat V_{\rm dis}+\hat V_\omega e^{-i\omega t}+\hat V_\omega^\dagger e^{i\omega t}$, where
\begin{equation}
\begin{aligned}
 \hat V_{\rm dis}(\mathbf r)=u\sum_{j=1}^{N_{\rm imp}}\delta^{(d)}(\mathbf r-\mathbf R_j),\quad \hat V_\omega=-e\hat{\mathbf r}\cdot\mathbf E(\omega).
\end{aligned}
\end{equation}
Here $\delta^{(d)}$ is the $d$-dimensional Dirac delta function,
$u$ is scalar real impurity strength, and $\hat{\mathbf r}$ is
the position operator.

To further illustrate the impurity average potential, the $N_{\rm imp}$ impurity positions are independently and
uniformly distributed in space of volume $V$ \cite{Mahan2000}. By absorbing the spatially uniform disorder average into the clean Hamiltonian, we could define $\delta\hat V_{\rm dis}=\hat V_{\rm dis}-\overline{\hat V_{\rm dis}}$. Following Refs.~\cite{Mahan2000, NagaosaEtAl2010}, for arbitrary states $\alpha,\beta,\lambda,\lambda'$, disorder averaging eliminates the first-order contribution, $\overline{\delta V_{\rm dis,\alpha\beta}}=0$, leaving only the second-order $\overline{\delta V_{\rm dis,\alpha\beta}\delta V_{\rm dis,\lambda\lambda'}}$ disorder average term.

\subsection{Origin of Non-Hermitian Interference Wilson-loop}
\begin{figure}
\centering
\includegraphics[width=1\linewidth]{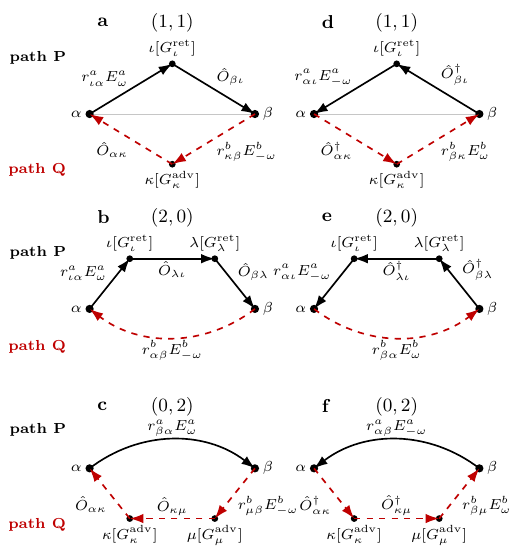}
\caption{{\bf Ordered scattering paths contributing to the second-order disorder induced photocurrents.} The left ({\bf a-c}) and right columns ({\bf d-f}) show contributions to forward ($W_{\alpha\to\beta}^{(\omega)}$) and reverse ($W_{\beta\to\alpha}^{(-\omega)}$) respectively. Solid black arrows denote the retarded path $P$, and dashed red arrows denote the advanced path $Q$. The pair $(l,s)$ gives the number of disorder scattering perturbations on $P$ and $Q$, respectively. Each branch contains one optical scattering perturbation, with its electric-dipole matrix element $r_{ij}^a=\langle i|\hat r^a|j\rangle$ and electric field $E_{\pm\omega}^a=E^a(\pm\omega)$ shown; here $E_{-\omega}^a=(E_\omega^a)^*$. In the left panels ({\bf a-c}), the retarded and advanced branches carry $E_\omega^a$ and $E_{-\omega}^b$, respectively; these frequency signs are reversed in the right panels ({\bf d-f}) The remaining scattering perturbations $\hat O_{ij}$ denote disorder perturbations.}
\label{fig:ordered_scattering_pair}
\end{figure}

So far, we have examined the non-Hermiticity of the field-free $T$-matrix operators and the constraints that anti-unitary symmetries impose on the transition rates. We now expand the anti-symmetric rates order by order in the impurity potential by resolving the scattering paths in the optical transition $\alpha\to\beta$. This analysis will clarify the origin of the field-free $T$-matrix's non-Hermiticity and reveal the Wilson-loop origin of $\mathcal Z_{\alpha\beta}^{ab}$ through interference between scattering paths, explaining how such interference produces non-reciprocal optical transition rates. We proceed by noticing that the successive terms in the $T$-matrix comprise scattering paths $P_{\alpha\beta}:\{|\alpha\rangle\to|\kappa\rangle\to\cdots\to|\beta\rangle\}$, where $|\kappa\rangle$ denotes an intermediate state along the path. This allows us to write the field-free transition amplitudes as $t_{\beta\alpha}^a(\omega)=\sum_P\boldsymbol{\mathcal M}_P^a\boldsymbol{\mathcal C}_P^{\rm ret}$, so that
\begin{equation}
 \boldsymbol{\mathcal M}_P^a
=\big[\prod_{P}O_{\zeta\kappa}\big]^a,
 \quad
 \boldsymbol{\mathcal C}_P^{\rm ret}
 =\prod_{\iota\in P_{\rm int}}
 [E_\iota-\epsilon_\iota+i\eta]^{-1},
 \label{eq:mandc}
\end{equation}
where the superscript $a$ denotes the polarization direction of the optical matrix element $r_{\zeta'\kappa'}^a=\langle\zeta'|\hat r^a|\kappa'\rangle$ along path $P$ as schematically shown in Fig.~\ref{fig:ordered_scattering_pair}. The scattering amplitude $O_{\zeta\kappa}=\langle\zeta|\hat O_{\zeta\kappa}|\kappa\rangle$ along legs  $\kappa\to\zeta$ in the path $P$, and $P_{\rm int}$ contains the intermediate states, see Fig.~\ref{fig:ordered_scattering_pair}. Here $E_\iota=\epsilon_\alpha+\sum_{\kappa\to\zeta\in P_{\alpha\iota}}\hbar\nu_{\zeta\kappa}$, where $\nu_{\zeta\kappa}$ is the frequency transfer along each leg from $\alpha\to\cdots\to\iota$; it equals $\omega$ for optical transition and zero at the static disorder scattering perturbations. When there are no intermediate states along $P$, we naturally set $\boldsymbol{\mathcal C}_P^{\rm ret}=1$. We note that $\boldsymbol{\mathcal C}_P^{\rm ret}$ is product of the retarded Green's functions. 

The decomposition in Eq.~\eqref{eq:mandc} expresses the squared transition amplitude compactly as
\begin{equation}
 |T_{\beta\alpha}^{(\omega)}|^2
 =e^2\sum_{P,Q}E^a(\omega)E^b(-\omega)
 \boldsymbol{\mathcal C}_{P,Q}\boldsymbol{\mathcal L}_{P,Q}^{ab},
\label{transition rate pathes expansion}
\end{equation}
where $\boldsymbol{\mathcal L}_{P,Q}^{ab}=\boldsymbol{\mathcal M}_P^a[\boldsymbol{\mathcal M}_Q^b]^*$ is a Wilson-loop tensor \cite{Sinitsyn2008, ShiSong2019, ShiEtAl2021,XiongShiSong2022} that describes the {\it complex}-valued interference amplitude of the optical transition. Its complete field-contracted form is $\boldsymbol{\mathcal L}_{P,Q}\equiv e^2E^a(\omega)E^b(-\omega)\boldsymbol{\mathcal L}_{P,Q}^{ab}=\boldsymbol{\mathcal M}_P\boldsymbol{\mathcal M}_Q^*$, where $\boldsymbol{\mathcal M}_P=-eE^a(\omega)\boldsymbol{\mathcal M}_P^a$. The propagator product is $\boldsymbol{\mathcal C}_{P,Q}=\boldsymbol{\mathcal C}_P^{\rm ret}\boldsymbol{\mathcal C}_Q^{\rm adv}$, see Fig.~\ref{fig:ordered_scattering_pair}, where $\boldsymbol{\mathcal C}_Q^{\rm adv}$ is the product of advanced Green's functions along path $Q$, obtained from the retarded product by replacing $i\eta\to-i\eta$.

The complex-valued nature of $E^a(\omega)E^b(-\omega)\boldsymbol{\mathcal L}_{P,Q}^{ab}$ produces path non-reciprocity. To see this, consider two paths with the same intermediate states but opposite orientations: a forward path $P_{\alpha\beta}:\{|\alpha\rangle\to|\kappa\rangle\to|\eta\rangle\to\cdots\to|\beta\rangle\}$ and a reverse path $P_{\beta\alpha}:\{|\alpha\rangle\leftarrow|\kappa\rangle\leftarrow|\eta\rangle\leftarrow\cdots\leftarrow|\beta\rangle\}$, as shown in Fig.~\ref{fig:ordered_scattering_pair}. For each leg along these reversed paths, the corresponding scattering perturbations are Hermitian conjugates; similarly, the intermediate energies $E_\iota$ remain the same on the energy-conserving shell. Notice that the reverse path optical transition transfers energy $-\hbar\omega$ since it corresponds to the electromagnetic field $E^a(-\omega)$. As a result, its energy-conserving delta function equals that of the forward process (see {\bf SM}) and we have 
\begin{equation}
 \boldsymbol{\mathcal C}^{\rm ret, \rm adv}(P_{\alpha\beta})
 =\boldsymbol{\mathcal C}^{\rm ret, \rm adv}(P_{\beta\alpha}).
 \label{eq:reverse-path-propagators}
\end{equation}
In contrast, the forward and reverse paths carry opposite Wilson-loop phases:
\begin{equation}
 \arg\Bigl[\boldsymbol{\mathcal L}_{\alpha\beta}\Bigr]
 =-\arg\Bigl[\boldsymbol{\mathcal L}_{\beta\alpha}\Bigr],
 \label{phase relation}
\end{equation}
reflecting the complex-valued nature of the Wilson-loop and linking geometric phases along scattering paths to non-reciprocal optical transitions.

Notice that in the lowest Born approximation, the field-contracted Wilson-loop factor $[L^{ab}(\omega)\operatorname{Re}\boldsymbol{\mathcal L}_{\alpha\beta}^{ab,(0)}-H^{ab}(\omega)\operatorname{Im}\boldsymbol{\mathcal L}_{\alpha\beta}^{ab,(0)}]=|E^a(\omega)\langle\beta|\hat r^a|\alpha\rangle|^2$ is real and nonnegative, with zero phase for a nonzero transition amplitude. This gives a reciprocal contribution when the conjugate photon branches are compared. However, at higher order in the scattering expansion, finite phases can develop due to the quantum geometry of the scattering states~\cite{Sinitsyn2008, MaAroraVignaleSong2023}. These geometric phases contribute to the non-Hermitian interference Wilson-loop through interference between scattering paths.

Using Eq.~\eqref{NH-WL}, we directly obtain the non-Hermitian interference Wilson-loop by summing across each path
\begin{align}
 \operatorname{Im}\mathcal Z_{\alpha\beta}^{ab,\omega}
 \hspace{-1mm}=&\frac12\sum_{P,Q}\Big\{
 \operatorname{Im}\boldsymbol{\mathcal L}_{P,Q}^{(ab)}
 \operatorname{Im}\boldsymbol{\mathcal C}_{P,Q}
 \hspace{-1mm}-\hspace{-1mm}\operatorname{Re}\!\left[
 \boldsymbol{\mathcal L}_{P,Q}^{[ab]}
 \boldsymbol{\mathcal C}_P^{\rm ret}
 \boldsymbol{\mathcal C}_Q^{\rm ret}
 \right]\Big\},
 \label{anti identity}\\
 \operatorname{Re}\mathcal Z_{\alpha\beta}^{ab,\omega}
 \hspace{-1mm}=&\frac12\sum_{P,Q}\Big\{
 \operatorname{Re}\boldsymbol{\mathcal L}_{P,Q}^{[ab]}
 \operatorname{Im}\boldsymbol{\mathcal C}_{P,Q}
 \hspace{-1mm}+\hspace{-1mm}\operatorname{Im}\!\left[
 \boldsymbol{\mathcal L}_{P,Q}^{(ab)}
 \boldsymbol{\mathcal C}_P^{\rm ret}
 \boldsymbol{\mathcal C}_Q^{\rm ret}
 \right]\Big\}.
 \label{eq:anti-identity-circular}
\end{align}
where $A^{(ab)}=(A^{ab}+A^{ba})/2$ and $A^{[ab]}=(A^{ab}-A^{ba})/2$ denote symmetrization and anti-symmetrization, respectively. Note that in Eq.~\eqref{anti identity}, the first (second) term is symmetric (antisymmetric) under $a\leftrightarrow b$, whereas in Eq.~\eqref{eq:anti-identity-circular}, the first (second) term is antisymmetric (symmetric). Thus, in obtaining the antisymmetric rates in Eq.~\eqref{eq:antirate}, the first term of Eq.~\eqref{anti identity} contributes to $[W^{A,(\omega)}]^\updownarrow$ under linearly polarized light, while the first term of Eq.~\eqref{eq:anti-identity-circular} contributes to $[W^{A,(\omega)}]^\circpol$ under circularly polarized light. Additionally, the second terms of Eqs.~\eqref{anti identity} and~\eqref{eq:anti-identity-circular} contribute to the circular and linear polarization components of $[R_{\alpha\beta}^\mu]^{S,(\omega)}$ through Eq.~\eqref{Rsymmetric}.

The scattering-path representations of $\mathcal Z_{\alpha\beta}^{ab}(\omega)$ in Eqs.~\eqref{anti identity} and~\eqref{eq:anti-identity-circular} demonstrate the multiple-scattering origin of the non-Hermitian interference Wilson-loop. In the Hermitian clean limit, $\mathcal Z_{\alpha\beta}^{ab}(\omega)$ vanishes. This can be seen term by term in Eqs.~\eqref{anti identity} and~\eqref{eq:anti-identity-circular}: only the direct optical path remains, with $\boldsymbol{\mathcal C}_0=1$, while $\boldsymbol{\mathcal L}_{0,0}^{(ab)}$ is purely real and $\boldsymbol{\mathcal L}_{0,0}^{[ab]}$ is purely imaginary. This demonstrates the essential role of multiple scattering in non-Hermitian optical transitions, whose leading contributions survive disorder averaging at order $u^2$.

\subsection{Disorder Induced Non-Hermitian Injection Current}
The previous scattering-path expansion of the antisymmetric rate provides a basis for calculating non-Hermitian photocurrents. As a warm-up, we consider the non-Hermitian injection current given by Eq.~\eqref{eq:jinjSA}. As discussed in Sections~$\mathbf I$ and $\mathbf {II}$, non-Hermitian photocurrents are governed by the antisymmetric transition rates $W_{\alpha\beta}^{A}$, whose linear and circular contributions involve $\operatorname{Im}\mathcal Z_{\alpha\beta}^{ab}$ and $\operatorname{Re}\mathcal Z_{\alpha\beta}^{ab}$, respectively. The path-by-path formulation in Eqs.~\eqref{anti identity} and~\eqref{eq:anti-identity-circular} allows us to organize the disorder corrections by the number of impurity perturbations. Let $t_{\omega;\beta\alpha}^{a,(l)}$ denote a field-free amplitude containing $l$ impurity perturbations and one optical scattering perturbation $\hat r^a$, with net energy transfer $\hbar\omega$. The pair $(l,s)$ contributes at order $u^{l+s}$ to the transition rate. Specifically, the $(0,0)$ pair gives the clean-limit response, whose antisymmetric part vanishes. The order-$u$ pairs $(1,0)$ and $(0,1)$ vanish upon averaging over centered disorder. Therefore, the leading disorder-averaged correction is of order $u^2$, with the field-free quadratic transition amplitude given by
\begin{equation}
\begin{aligned}[b]
 \overline{\mathcal K_{\omega;\beta\alpha}^{ab,(2)}}
 =\overline{
 t_{\omega;\beta\alpha}^{a,(1)}
 \big(t_{\omega;\beta\alpha}^{b,(1)}\big)^*}+\overline{
 t_{\omega;\beta\alpha}^{a,(2)}
 \big(t_{\omega;\beta\alpha}^{b,(0)}\big)^*}
 +\overline{
 t_{\omega;\beta\alpha}^{a,(0)}
 \big(t_{\omega;\beta\alpha}^{b,(2)}\big)^*}.
\end{aligned}
\label{eq:injection-kernel-u2}
\end{equation}
These terms correspond to the $(1,1)$, $(2,0)$, and $(0,2)$ path pairs respectively, see Fig.~\ref{fig:ordered_scattering_pair}. The first contains one impurity perturbation in each amplitude, while the latter two describe interference between a clean amplitude and an amplitude containing two impurity perturbations. Each amplitude includes all allowed orderings of the optical and impurity scattering perturbations.

Using the polarization decomposition in Eq.~\eqref{eq:electric bilinear detachment}, the corresponding transition rate is
\begin{equation}
\begin{aligned}[b]
 \overline{W_{\alpha\to\beta}^{(2)}}
 =\frac{2\pi e^2}{\hbar}\sum_{\pm\omega}\Delta_{\omega}
 \Big[
 L^{ab}_\omega
 \operatorname{Re}\overline{\mathcal K_{\omega;\beta\alpha}^{ab,(2)}}-H^{ab}_\omega
 \operatorname{Im}\overline{\mathcal K_{\omega;\beta\alpha}^{ab,(2)}}
 \Big].
\end{aligned}
\label{eq:injection-rate-u2}
\end{equation}
These two terms give the linear and circular contributions, respectively. Their symmetric and antisymmetric parts follow from $\overline{W_{\alpha\beta}^{S,A,(2)}}=(\overline{W_{\alpha\to\beta}^{(2)}}\pm\overline{W_{\beta\to\alpha}^{(2)}})/2$, where the reverse rate includes the conjugate photon branch. Substituting these rates into the injection-current expressions gives the Hermitian and non-Hermitian contributions to injection photocurrents.

The disorder-induced linear (polarized) injection current is the ballistic photocurrent associated with asymmetric photoexcited carrier generation \cite{BelinicherSturman1988,BelinicherSturman1980,LeppenenGolub2023,Sturman2020,SturmanFridkin1992,DaiEtAl2021, Ivchenko, DaiRappe2021}. 
Here we also identify the circularly polarized non-Hermitian interband injection contribution. 
It provides a circular photocurrent beyond the clean-limit injection response and can survive in $\mathcal{PT}$-symmetric magnets (unlike that of the clean symmetric circular injection photocurrent). The disorder averages follow standard procedures but involve lengthy intermediate expressions.

\subsection{Disorder Induced Non-Hermitian Shift Currents}

With the path-by-path transition-rate expansion constructed above, we now turn to the non-Hermitian shift currents. Eqs.~\eqref{eq:shiftcurrent1} and~\eqref{eq:shiftcurrent2} describe the full shift currents compactly, but the disorder average must be taken on the combined product $W_{\alpha\beta}^{(\omega)}R_{\alpha\beta}^{\mu,(\omega)}$. This is difficult in practice because the shift vector contains the phase $\operatorname{arg}T_{\beta\alpha}^{(\omega)}$. Simply averaging the rate and displacement separately would discard their disorder correlations. For practical evaluation, we rewrite $\Phi_{\alpha\beta}^\mu(\omega) \equiv W_{\alpha\beta}^{(\omega)}R_{\alpha\beta}^{\mu,(\omega)} $ for each frequency process and by recalling the identity $\operatorname{Im}[(T_{\beta\alpha})^*\mathcal D_\mu^{\alpha\beta} T_{\beta\alpha}]=-|T_{\beta\alpha}|^2R^\mu_{\alpha \to \beta}$, we write $\Phi_{\alpha\beta}^\mu(\omega)$ in terms of $t_{\omega; \beta \alpha}^a$ as
\begin{equation}
\begin{aligned}[b]
 \Phi_{\alpha\beta}^\mu(\omega) = -\frac{2\pi e^2}{\hbar}
&\Delta_\omega\Big\{L^{ab}(\omega)\operatorname{Im}\!\left[
 (t_{\beta\alpha}^{b})^*
 \mathcal D_\mu^{\alpha\beta}t_{\beta\alpha}^{a}
 \right]\\
 &+H^{ab}(\omega)\operatorname{Re}\!\left[
 (t_{\beta\alpha}^{b})^*
 \mathcal D_\mu^{\alpha\beta}t_{\beta\alpha}^{a}
 \right]\Big\},
\end{aligned}
\label{eq:iden}
\end{equation}
with the covariant derivative $\mathcal D_\mu^{\alpha\beta}=\partial_\alpha^\mu+\partial_\beta^\mu-i(\mathcal A_\beta^\mu-\mathcal A_\alpha^\mu)$. Both absorption ($+\omega$) and emission ($-\omega$) processes are included when constructing the shift current, $j_{\rm shift}^\mu=e\sum_{\alpha,\beta,\pm\omega}f_\alpha(1-f_\beta)\Phi_{\alpha\beta}^\mu(\omega)$. Since the external field is independent of the electronic momenta, the momentum derivatives act only on the field-free transition amplitudes. This $\Phi$ formulation allows the disorder average to be evaluated path-by-path.

For each frequency channel, the combinations entering Eqs.~\eqref{eq:shiftcurrent1} and~\eqref{eq:shiftcurrent2} satisfy
\begin{align}
 W^{S,(\omega)}_{\alpha\beta}R^{A,(\omega),\mu}_{\alpha\beta}
 +W^{A,(\omega)}_{\alpha\beta}R^{S,(\omega),\mu}_{\alpha\beta}
 =\frac{\Phi_{\alpha\beta}^\mu(\omega)-\Phi_{\beta\alpha}^\mu(-\omega)}{2},
 \label{eq:shift-kernel-exchange-odd}
\end{align}
\begin{align}
 W^{S,(\omega)}_{\alpha\beta}R^{S,(\omega),\mu}_{\alpha\beta}
 +W^{A,(\omega)}_{\alpha\beta}R^{A,(\omega),\mu}_{\alpha\beta}
 =\frac{\Phi_{\alpha\beta}^\mu(\omega)+\Phi_{\beta\alpha}^\mu(-\omega)}{2},
 \label{eq:shift-kernel-exchange}
\end{align}
where the Hermitian and non-Hermitian shift currents select the odd and even parts of $\Phi_{\alpha\beta}^\mu(\omega)$, respectively. The non-Hermitian shift current receives contributions from a symmetric shift weighted by a reciprocal (symmetric) rate and an antisymmetric shift weighted by a antisymmetric rate. Both of these contributions to the non-Hermitian shift current vanish in the clean limit where $R^S_{\alpha\beta}=W^A_{\alpha\beta}=0$. Multiple-scattering interference activates these terms, producing an exchange-even $\Phi_{\alpha\beta}^\mu$ for the non-Hermitian shift current.

To resolve $\Phi_{\alpha\beta}^\mu(\omega)$ into scattering paths, write $t_{\beta\alpha}^{a}(\omega)=\sum_P\boldsymbol{\mathcal C}_P\boldsymbol{\mathcal M}_P^a$, as defined in Section~{\bf IV A}. The conjugate amplitude supplies the advanced path $Q$. Since $\boldsymbol{\mathcal C}_P$ is gauge invariant, the product rule gives $\mathcal D_\mu^{\alpha\beta}(\boldsymbol{\mathcal C}_P\boldsymbol{\mathcal M}_P^a)=\boldsymbol{\mathcal C}_P\mathcal D_\mu^{\alpha\beta}\boldsymbol{\mathcal M}_P^a+(\partial_\mu\boldsymbol{\mathcal C}_P)\boldsymbol{\mathcal M}_P^a$, where $\partial_\mu=\partial_\alpha^\mu+\partial_\beta^\mu$. This separates derivatives of the numerator matrix elements from those of the propagator denominators. Hence, we obtain
\begin{equation}
 t_{\beta\alpha}^{b,*}
 \mathcal D_\mu^{\alpha\beta}t_{\beta\alpha}^{a}
 =\sum_{P,Q}\boldsymbol{\mathcal C}_{P,Q}
 \boldsymbol{\mathcal M}_Q^{b,*}\mathcal D_\mu^{\alpha\beta}\boldsymbol{\mathcal M}_P^a+\boldsymbol{\mathcal C}_Q^*(\partial_\mu\boldsymbol{\mathcal C}_P)
 \boldsymbol{\mathcal L}_{P,Q}^{ab},
\label{eq:path-shift-kernel}
\end{equation}
Here $\boldsymbol{\mathcal C}_P=\boldsymbol{\mathcal C}_P^{\rm ret}$, so that $\boldsymbol{\mathcal C}_Q^*=\boldsymbol{\mathcal C}_Q^{\rm adv}$. 

Using Eq.~\eqref{eq:shift-kernel-exchange-odd} and substituting Eq.~\eqref{eq:path-shift-kernel} into Eq.~\eqref{eq:iden}, the Hermitian shift current in Eq.~\eqref{eq:shiftcurrent1} follows from the exchange-odd part of $\Phi_{\alpha\beta}^\mu$:
\begin{widetext}
\begin{equation}
 \mathbf j_{\rm shift}^{{\rm H},\mu}
 =\sum_{\substack{\alpha,\beta,\omega\\P,Q}}\tilde F_{\alpha\beta}^{\omega}
 \Big\{L^{ab}(\omega)\overline{\operatorname{Re}\boldsymbol{\mathcal C}_{P,Q}\operatorname{Im}\boldsymbol{\mathcal L}_{P,Q}^{ab;\mu}+\operatorname{Im}\boldsymbol{\mathcal L}_{P,Q}^{ab}\operatorname{Re}\boldsymbol{\mathcal C}_{P,Q}^{\mu}}
 +H^{ab}(\omega)\overline{\operatorname{Re}\boldsymbol{\mathcal C}_{P,Q}\operatorname{Re}\boldsymbol{\mathcal L}_{P,Q}^{ab;\mu}+\operatorname{Re}\boldsymbol{\mathcal L}_{P,Q}^{ab}\operatorname{Re}\boldsymbol{\mathcal C}_{P,Q}^{\mu}}\Big\}.
\label{eq:symshift-new}
\end{equation}
Here $\tilde F_{\alpha\beta}^{\omega}=-(\pi e^3/\hbar)\Delta_\omega f_{\alpha\beta}$, and we define $\boldsymbol{\mathcal L}_{P,Q}^{ab;\mu}=(\boldsymbol{\mathcal M}_Q^b)^*\mathcal D_\mu^{\alpha\beta}\boldsymbol{\mathcal M}_P^a$ and $\boldsymbol{\mathcal C}_{P,Q}^{\mu}=\boldsymbol{\mathcal C}_Q^*\partial_\mu\boldsymbol{\mathcal C}_P$. The corresponding field-contracted numerator derivative is $\boldsymbol{\mathcal L}_{P,Q}^{\mu}=e^2E^a(\omega)E^b(-\omega)\boldsymbol{\mathcal L}_{P,Q}^{ab;\mu}=\boldsymbol{\mathcal M}_Q^*\mathcal D_\mu^{\alpha\beta}\boldsymbol{\mathcal M}_P$, where $\boldsymbol{\mathcal M}_P=-eE^a(\omega)\boldsymbol{\mathcal M}_P^a$. The overbar denotes the disorder average of each complete path product. In the clean limit, there is only a single optical path with $\boldsymbol{\mathcal C}_0=1$ and $\boldsymbol{\mathcal M}_0^a=r_{\beta\alpha}^a$~\cite{AhnGuoNagaosa2020}. Thus, only the terms proportional to $\operatorname{Re}\boldsymbol{\mathcal C}_{P,Q}$ in Eq.~\eqref{eq:symshift-new} survive. Using $\boldsymbol{\mathcal L}_{0,0}^{ab;\mu}=(r_{\beta\alpha}^b)^*\mathcal D_\mu^{\alpha\beta}r_{\beta\alpha}^a$ and $[V_\omega]_{\beta\alpha}=-eE^a(\omega)r_{\beta\alpha}^a$, the contracted derivative reduces to $\boldsymbol{\mathcal L}_{0,0}^{\mu}=[V_\omega]_{\beta\alpha}^*\mathcal D_\mu^{\alpha\beta}[V_\omega]_{\beta\alpha}$, whose imaginary part is $-|[V_\omega]_{\beta\alpha}|^2R_{\alpha\to\beta}^{\mu,(0)}(\omega)$. Substitution into Eq.~\eqref{eq:symshift-new} gives the familiar clean shift current $\mathbf j_{\rm shift}^{\mu}=\frac{\pi e}{\hbar}\sum_{\alpha,\beta,\pm\omega}f_{\alpha\beta}\Delta_\omega|[V_\omega]_{\beta\alpha}|^2R_{\alpha\to\beta}^{\mu,(0)}(\omega)$. Multiple scattering modifies these numerator-derivative terms and can also activate the terms containing $\operatorname{Re}\boldsymbol{\mathcal C}_{P,Q}^{\mu}$, which describe the momentum dependence of the intermediate Green's functions.

Similarly, using Eq.~\eqref{eq:shift-kernel-exchange}, the exchange-even part of $\Phi_{\alpha\beta}^\mu$ gives the non-Hermitian shift current in Eq.~\eqref{eq:shiftcurrent2}:
\begin{equation}
 \mathbf j_{\rm shift}^{{\rm NH},\mu}
 =\sum_{\substack{\alpha,\beta,\omega\\P,Q}}\tilde F_{\alpha\beta}^{\omega}\widetilde N_{\beta\alpha}
 \Big\{L^{ab}(\omega)\overline{\operatorname{Im}\boldsymbol{\mathcal C}_{P,Q}\operatorname{Re}\boldsymbol{\mathcal L}_{P,Q}^{ab;\mu}+\operatorname{Re}\boldsymbol{\mathcal L}_{P,Q}^{ab}\operatorname{Im}\boldsymbol{\mathcal C}_{P,Q}^{\mu}}
 -H^{ab}(\omega)\overline{\operatorname{Im}\boldsymbol{\mathcal C}_{P,Q}\operatorname{Im}\boldsymbol{\mathcal L}_{P,Q}^{ab;\mu}+\operatorname{Im}\boldsymbol{\mathcal L}_{P,Q}^{ab}\operatorname{Im}\boldsymbol{\mathcal C}_{P,Q}^{\mu}}\Big\}.
\label{eq:skewshift-new}
\end{equation}
Both linear and circular polarization contributions to the non-Hermitian shift photocurrent involve imaginary parts of the propagator products or their momentum derivatives. All terms in Eq.~\eqref{eq:skewshift-new} vanish for the direct optical transition in the clean limit, since $\operatorname{Im}\boldsymbol{\mathcal C}_{0,0}=0$ and $\boldsymbol{\mathcal C}_{0,0}^{\mu}=0$. They can become finite through disorder-assisted scattering. For centered local disorder, the leading disorder-averaged corrections can arise at order $u^2$ from the $(1,1)$, $(2,0)$, and $(0,2)$ path pairs shown in Fig.~\ref{fig:ordered_scattering_pair}.
\end{widetext}

We next apply this formulation to bound excitons, focusing on circularly polarized excitation in a non-magnetic system. This setting allows us to examine how a disorder-assisted scattering process produces a helicity-dependent real-space charge displacement during exciton creation which persists even in non-magnetic systems; as we will see, these arise from non-Hermitian excitonic circular shift photocurrents. Similarly, we will also examine disorder-assisted linear excitonic transition shift currents in high symmetry systems; in which we will see how disorder can also unblock Hermitian excitonic linear shift currents.

\begin{figure*}
    \centering
\includegraphics[width=1\linewidth]{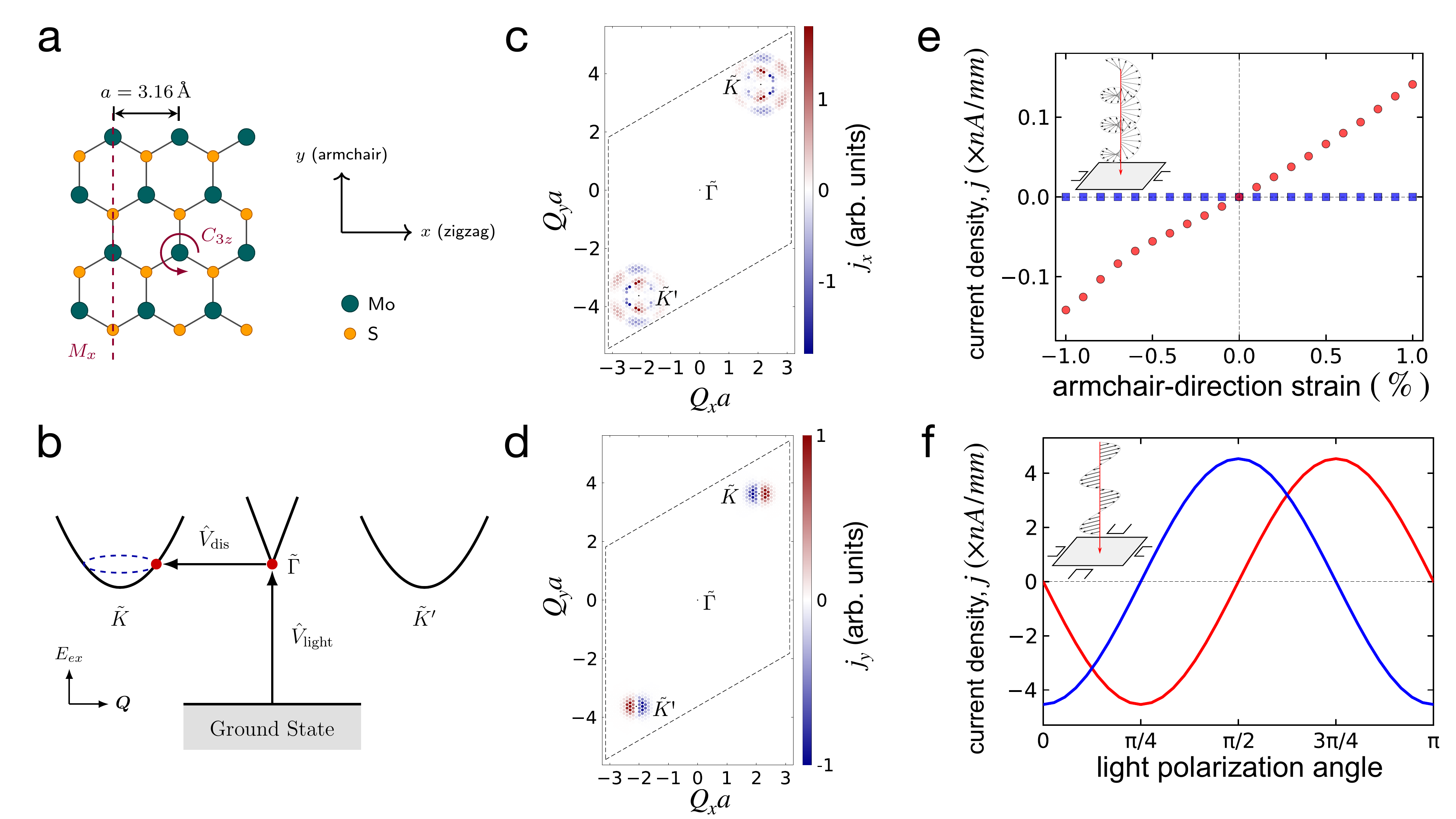}
    \caption{{\bf Disorder-assisted circular non-Hermitian excitonic shift photocurrents and and disorder-assisted linear Hermitian excitonic shift currents in monolayer MoS$_2$.} {\bf a.} Lattice structure of monolayer MoS$_2$. Teal/orange dots denote Mo/S atoms. Pristine MoS$_2$ exhibits $C_{3v}$ point-group symmetry, generated by a mirror reflection $M_x$ and a three-fold rotation $C_{3z}$. Strain along the armchair direction breaks $C_{3z}$ symmetry while preserving $M_x$. {\bf b.} Schematic of excitonic band minima as a function of center-of-mass momentum $Q$ in MoS$_2$ and the disorder-assisted optical excitation of finite-$\bm{Q}$ excitons. Exchange interactions produce a Dirac-like dispersion near $\bm{Q}=\tilde{\Gamma}$\cite{yu2014dirac}. The lower-energy exciton minima Excitons near $\bm{Q}=\tilde{K}$ and $\bm{Q}=\tilde{K}'$ enable elastic scattering of optically excited $\tilde{\Gamma}$ exciton in their vicinity, providing the dominant contributions to the shift currents. {\bf c.} Momentum-resolved contributions to the circular non-Hermitian shift current in monolayer MoS$_2$ under $0.1\%$ strain. The dominant contributions arise from finite-$\bm{Q}$ excitons near $\tilde{K}$ and $\tilde{K}'$. {\bf d.} Momentum-resolved contributions to the linear shift current in unstrained monolayer MoS$_2$ with light polarization along $x$ direction. The dominant contributions likewise arise from finite-$\bm{Q}$ excitons near $\tilde{K},\tilde{K}'$. The results in panels ({\bf c-d}) are calculated on an $60 \times 60$ lattice. {\bf e.} Excitonic non-Hermitian circular shift current as a function of varying strain in monolayer MoS$_2$. Uniaxial strain is applied along armchair ($y$) direction. Red circles (blue squares) denote $j_x$ ($j_y$). The applied strain breaks $C_{3z}$ symmetry, allowing for a finite circular shift current. $M_x$ forces $j_y$ to vanish. {\bf f.} Disorder-assisted excitonic linear shift current as a function of light polarization in unstrained MoS$_2$. Red (blue) solid lines denote $j_x$ ($j_y$). Disorder-assisted optical excitation of finite-$\bm{Q}$ excitons generate a non-zero linear shift current. Notice that such linear excitonic shift currents vanishes in the clean limit in high symmetry MoS$_2$. The results in panels ({\bf e-f}) converge on a $96\times 96$ lattice. } 
    \label{fig:fig3}
\end{figure*}

\section{Numerical Illustration and Disorder Induced Excitonic Shift Currents in MoS$_2$}\label{sec:V}

{We now turn to illustrate non-Hermitian photocurrents and disorder corrections to shift photocurrents in a model system. As we argue below, excitonic transition photocurrents are a particularly interesting platform since their injection and ballistic vanish identically due to the bound nature of their electron-hole pairs.}  

To see this, we first note that excitons are bound states of particle-hole pairs due to Coulomb interactions. They can be described by coherent superpositions of particle-hole excitations from the ground state: $\ket{n_{ex,\bm{Q}}}=\sum\limits_{\textbf{k}}\psi_{n_{ex},\bm{Q}}(\textbf{k})c^{\dagger}_{c,\textbf{k}+\bm{Q}}c_{v,\textbf{k}}\ket{0}$, with $\ket{0}$ the insulating ground state with filled valence bands. Here, $n$ labels the exciton state, $\bm{Q}$ is its center-of-mass momentum, and $\psi_{n_{ex},\bm{Q}}(\textbf{k})$ the exciton envelope function with $\textbf{k}$ the relative momentum. $c_{c/v,\textbf{k}}^{\dagger}$ creates a conduction/valence band Bloch state at $\textbf{k}$. The bound nature of excitons is reflected in the localization of their envelope function in the relative coordinate with exponential decay $\psi_{n_{ex},\bm{Q}}(\textbf{r})\sim e^{-|\textbf{r}|/\xi}$. For a bound exciton eigenstate, the average relative velocity of the electron and hole vanishes, yielding zero net charge current, $-e\braket{\hat{v}_e-\hat{v}_h}=0$~\cite{Wannier1937, YangSrivastavaSong2026}. Consequently, bound excitons do not exhibit an injection current and hence provide a particularly clean platform to demonstrate shift-current phenomena. In the following, we demonstrate how disorder induces shift currents that vanish in the clean limit, using linear and circular excitonic shift currents as examples. 

\subsection{Circular Excitonic non-Hermitian Shift Current}
Our first example is the circular non-Hermitian shift current in a $\mathcal{T}$-symmetric system, exemplified by monolayer MoS$_2$ with $C_{3v}$ point group symmetry. See Fig.~\ref{fig:fig3}(a) for illustration of the lattice and its crystalline symmetries. In the clean limit, only $\bm{Q}=\bm{0}$ excitons are excited by light-matter interaction, and the circular excitonic shift current vanishes due to time-reversal symmetry. Disorder scattering, however, enables non-Hermitian indirect transitions to finite-$\bm{Q}$ excitons, giving rise to a non-zero circular excitonic shift current despite time-reversal symmetry. The exciton dispersion of MoS$_2$ exhibits three local minima at $\bm{Q}=\tilde{\Gamma},\tilde{K},\tilde{K}'$. All possible disorder-assisted scattering pathways are shown in Fig.~\ref{fig:ordered_scattering_pair} to $u^2$. To identify the dominant pathways, we note that light-matter-interaction conserves the total momentum, and disorder scattering is elastic. First, due to the large ($\sim$ eV) energy difference between the ground state and exciton states, the ground state can only be excited to a $\tilde{\Gamma}$ exciton via light-matter interaction. And $\tilde{\Gamma}$ exciton can be scattered by disorder into finite $\bm{Q}$ excitons that are nearly on-resonant, with large momentum transfer allowed. Consequently, the dominant $(1,1)$ pathways involve optical excitation of $\tilde{\Gamma}$ excitons from the ground state, followed by disorder scattering into finite-$\bm{Q}$ near $\tilde{K}$ and $\tilde{K}'$ excitons having energies close to $\tilde{\Gamma}$ excitons. See Fig.~\ref{fig:fig3}(b) for an illustration of exciton band minima and the dominant $(1,1)$ pathways to finite-$\bm{Q}$ excitons. As we show below, $(0,2)$ and $(2,0)$ processes do not contribute to the total current in the systems we consider.

As excitons are coherent superpositions of particle-hole excitations over relative momenta, their shift current response simplifies considerably. In particular, exciton energies are insensitive to momentum variations, yielding $\boldsymbol{\mathcal{C}}_{P,Q}^{\mu}=0$, while the derivative of scattering phase shift $\partial_{\textbf{k}}\text{arg}(T)$ in Eq.~\eqref{eq:shift} vanishes, allowing for the factorization $\boldsymbol{\mathcal{L}}_{P,Q}^{\mu}=iR_{0\rightarrow ex}^{\mu}\boldsymbol{\mathcal{L}}_{P,Q}$. Here $R_{0\rightarrow ex}^{\mu}$ is the excitonic transition shift vector determined by the exciton envelope function and the conduction/valence band Bloch functions\cite{YangSrivastavaSong2026}. Here $R_{0\rightarrow ex}^{\mu}$ is the excitonic transition shift vector determined by the exciton envelope function and the conduction/valence band Bloch functions\cite{YangSrivastavaSong2026}. They are given by: 
\begin{align}
R^{\mu}_{0\rightarrow n_{ex,\bm{Q}}}=\sum\limits_{\textbf{r},\textbf{r}'}\rho(\textbf{r},\textbf{r}')[\textbf{r}\delta_{\textbf{r},\textbf{r}'}+d_{cv}(\Delta_{\textbf{r}})].
\end{align}
Here $\rho(\textbf{r},\textbf{r}')=\psi_{n_{ex},\bm{Q}}^*(\textbf{r})\psi_{n_{ex},\bm{Q}}(\textbf{r}')$ is the reduced density matrix, $\Delta_{\textbf{r}}=\textbf{r}'-\textbf{r}$, $d_{cv}(\textbf{R})\equiv e^{i\bm{Q}\cdot \textbf{R}}d_c(\textbf{R})-d_v(\textbf{R})$ and $d_{c/v}(\textbf{R})=\sum\limits_{\textbf{k}}e^{i\textbf{k}\cdot\textbf{R}}\mathcal{A}_{c/v}(\textbf{k})$ is the Fourier transformed Berry connection.

Consequently, the circular excitonic non-Hermitian shift conductivity in Eq.~\eqref{eq:skewshift-new} reduces to:
\begin{align}
&\sigma^{\text{NH}}_{\mu;xy}\!=\!\sum\limits_{\!m,\bm{Q},n,p\!}\! \frac{\pi e^3}{\hbar}\Delta_{\omega}R^{\mu}_{0\rightarrow m_{ex,\bm{Q}}\!}\operatorname{Re}(\boldsymbol{\mathcal{L}}^{xy}_{\!P_p,Q_n\!})\operatorname{Im}(\boldsymbol{\mathcal{C}}_{\!P_p,Q_n\!}).\label{eq:skew_conductivity_exciton}
\end{align}
Pathways are $P_p:\ket{0}\rightarrow \ket{p_{ex,\bm{0}}}\rightarrow \ket{m_{ex,\bm{Q}}}$ and $Q_n:\ket{0}\rightarrow \ket{n_{ex,\bm{0}}}\rightarrow \ket{m_{ex,\bm{Q}}}$, with summations over exciton states $m,n,p$ and final state momentum $\bm{Q}$. The corresponding quantities are $\boldsymbol{\mathcal{L}}^{xy}_{P_p,Q_n}=r^x_{0,n_{ex,\bm{0}}}\overline{\bra{n_{ex,\bm{0}}}\hat{V}_{\text{dis}}\ket{m_{ex,\bm{Q}}}\bra{m_{ex,\bm{Q}}}\hat{V}_{\text{dis}}\ket{p_{ex,\bm{0}}}}r^y_{p_{ex,\bm{0}},0}$, and $\boldsymbol{\mathcal{C}}_{P_p,Q_n}=(\varepsilon_{m_{ex},\bm{Q}}-\varepsilon_{n_{ex},\bm{0}}-i\eta)^{-1}(\varepsilon_{m_{ex},\bm{Q}}-\varepsilon_{p_{ex},\bm{0}}+i\eta)^{-1}$. The disorder average is evaluated over Bloch state matrix elements, followed by summation weighted by excitonic envelope functions. We focus on zero temperature and therefore set both $f$ and $\tilde{N}$ to be 1 hereafter. We only include the $(1,1)$ contribution in Eq.~\eqref{eq:skew_conductivity_exciton}. The $(0,2)/(2,0)$ processes do not contribute to the total response because they are proportional to the net circular absorption rate of the $\tilde{\Gamma}$ exciton, which vanishes under $M_x$ symmetry.

Interestingly, because the bound nature of the excitons render excitonic transition shift vectors independent of the polarization of light~\cite{YangSrivastavaSong2026}, the excitonic non-Hermitian circular shift current can be understood in a simple physical way. Each $\bm{Q}$ indirect transition possess its own shift vector while being pumped by non-Hermitian optical transitions. As we will see below, the finite $\bm{Q}$ indirect transitions have imbalanced transition rates that discriminate between LH and RH circularly polarized light allowing to pump a finite shift current. Notice that non-Hermiticity is essential here since the (Hermitian) symmetric rates are odd under time-reversal.

To illustrate the excitonic non-Hermitian circular shift current in Eq.~\eqref{eq:skew_conductivity_exciton}, we performed numerical simulations using a three-band tight-binding model of monolayer MoS$_2$ with a screened Keldysh potential and solve the Bethe-Salpeter equation including both the direct and exchange interactions to obtain the exciton eigen-states and eigen-energies~\cite{Keldysh1979,cudazzo2011dielectric}. The tight-binding parameters and interaction parameters are taken from Refs.~\cite{liu2013three,wu2015exciton}. We use a disorder strength $n_{\rm imp}u^2=0.5\, \text{eV}a^2$, where  $a=3.16\,$\AA\, is the lattice constant, and an electric field amplitude of $E=300\rm{V}\,\rm{cm}^{-1}$. We set the photon energy to be on resonant with the lowest energy $\tilde{\Gamma}$ exciton and compute the response. The energy delta function is regularized by a Gaussian $\delta_{\gamma}(E)=\frac{1}{\gamma\sqrt{2\pi}}e^{-\frac{E^2}{2\gamma^2}}$ with width $\gamma=50\,\rm{meV}$. As we evaluate the response on a finite momentum grid of $\bm{Q}$, we use a finite broadening $\eta=25\,\rm{meV}$ in the $T$-matrix to ensure numerical convergence with system size.

As pristine MoS$_2$ has $C_{3z}$ symmetry which forbids a bulk circular photogalvanic current, we apply strain along the armchair ($y$) direction following Ref.~\cite{rostami2015theory}. The momentum-resolved contributions to $j_x$ at strain $0.1\%$ in Fig.~\ref{fig:fig3}(c) show that excitons near $\tilde{K},\tilde{K}'$ that are resonant with the $\tilde{\Gamma}$ exciton dominate the circular non-Hermitian shift current. Note that contributions from time-reversal-related momenta $\bm{Q}$ are identical. The resulting current as a function of strain is shown in Fig.~\ref{fig:fig3}(e). The residual mirror symmetry $M_x:x\rightarrow -x$ in the strained system forces $j_y$ (blue solid dots) to vanish, while $j_x$ (red solid dots) remains finite, demonstrating a nonzero disorder-induced non-Hermitian circular shift current.

This is particularly surprising for several reasons. First, in the clean limit, circular shift photocurrents have long been thought to vanish making circular injection photocurrents the prime mechanism for circular photogalvanic currents in non-magnetic materials~\cite{DeJuan, AhnGuoNagaosa2020, MaEtAl2023}. Second, clean injection currents vanish for excitons. These two facts make the non-Hermitian circular shift photocurrent we find here particularly non-intuitive: not only does it not vanish in a non-magnetic system, it dominates the circular photogalvanic effect for excitonic transitions.

\subsection{Disorder-induced Linear Excitonic Shift Current in $C_3$ Symmetric Material}
Our second example is the disorder-induced linear shift current in a $\mathcal{T}$-symmetric system with non-polar point group symmetry. Although linear shift currents are generally allowed by time-reversal symmetry, they can be forbidden by additional crystalline symmetries. In the clean limit, vertical transitions excite only center-of-mass $\bm{Q}=\bm{0}$ excitons (the $\tilde{\Gamma}$ excitons). Since $C_{3z}$ symmetry requires $\hat{U}_{C_{3z}}\bm{R}(\bm{0})=\bm{R}(\bm{0})$, the in-plane shift vector satisfies $\bm{R}(\bm{0})=\bm{0}$, forcing the linear excitonic shift current to vanish despite being allowed by time-reversal symmetry~\cite{YangSrivastavaSong2026}.

The disorder-induced shift current, however, is not necessarily subject to the same constraint. Disorder scattering enables optical excitation of finite-$\bm{Q}$ excitons carrying non-zero shift vectors. Although the excitonic shift vectors at $C_{3z}$ related $\bm{Q}$ momenta are related by $C_{3z}$ symmetry and add up to zero, under linearly polarized light, these excitons contribute to the shift current with unbalanced weights, allowing for a finite net linear shift current. The corresponding conductivity is given by:
\begin{align}
&\sigma^{\text{H}}_{\mu;ab}\!=\!\sum\limits_{\!m,\bm{Q},n,p\!} \frac{\pi e^3}{\hbar}\Delta_{\omega}R^{\mu}_{0\rightarrow m_{ex,\bm{Q}}\!}\operatorname{Re}(\boldsymbol{\mathcal{L}}^{ab}_{\!P_p,Q_n\!})\operatorname{Re}(\boldsymbol{\mathcal{C}}_{\!P_p,Q_n\!}), 
\end{align}
which is obtained by applying the simplification $\boldsymbol{\mathcal{C}}^{\mu}=0$ and $\boldsymbol{\mathcal{L}}^{\mu}=iR^{\mu}\boldsymbol{\mathcal{L}}$ to Eq.~\eqref{eq:symshift-new} valid for excitonic transitions. Here the dominant scattering pathways are the same as in the circular non-Hermitian shift current, see Fig.~\ref{fig:fig3}, and $\boldsymbol{\mathcal{C}}_{P_p,Q_n}$ is similarly defined as before, and $\boldsymbol{\mathcal{L}}^{ab}_{P_p,Q_n}=r^{a}_{0,n_{ex,\bm{0}}}\overline{\bra{n_{ex,\bm{0}}}\hat{V}_{\text{dis}}\ket{m_{ex,\bm{Q}}}\bra{m_{ex,\bm{Q}}}\hat{V}_{\text{dis}}\ket{p_{ex,\bm{0}}}}r^{b}_{p_{ex,\bm{0}},0}$. Here we only include the $(1,1)$ contribution, because only $\tilde{\Gamma}$ excitons are optically excited in $(0,2)$ and $(2,0)$ processes, and their excitonic transition shift vectors vanish by $C_{3z}$ symmetry. The momentum-resolved contributions to $j_y$ in Fig.~\ref{fig:fig3}(d) under a light with polarization along $x$ show that excitons near $\tilde{K}$ and $\tilde{K}'$ resonant with the $\tilde{\Gamma}$ exciton dominate the linear shift current. The resulting net current flows along $y$, consistent with a symmetry allowed finite $\sigma^{yxx}$. The total linear shift current (red/blue solid lines for $j_x/j_y$) as a function of light polarization angle is shown in Fig.~\ref{fig:fig3}(f), demonstrating a disorder-induced linearly polarized excitonic shift current that are otherwise forbidden by the non-polar point group in the clean limit.

The emergence of a finite linear excitonic shift current in unstrained MoS$_2$ is surprising: $C_{3z}$ symmetry forbids the clean-limit response, and yet disorder activates it despite preserving this symmetry on average. By enabling transition to finite-$\bm{Q}$ excitons whose shift vectors not vanish, disorder provides a route to excitonic photocurrents in nonpolar materials distinct from strain-induced symmetry lowering~\cite{YangSrivastavaSong2026}. Because the $\bm{Q} = 0$ excitonic shift photocurrent as well as injection/ballistic excitonic currents vanish, such disorder-assisted excitonic shift currents provide the dominant photocurrent for excitonic transitions.  This mirrors the as-yet-unexplained finite and polarization-dependence of the excitonic shift current experimentally observed in {\it unstrained} 3R-MoS$_2$~\cite{DongEtAl2023}. This correspondence provides evidence for the disorder-assisted linear excitonic shift current. 

\section{Summary and Discussion}

In this work, we have systematically demonstrated the impact of indirect non-Hermitian optical transitions in nonlinear photocurrent response. We have introduced a new geometric photocurrent: a non-Hermitian shift current. This depends on unconventional geometric phases, specifically the non-Hermitian interference Wilson-loop $\mathcal{Z}_{\alpha \beta}$ that vanishes in the Hermitian limit, e.g., for the simple direct electric-dipole transitions~\cite{SipeShkrebtii2000,AhnEtAl2022}. It is important to notice that this non-Hermiticity in optical transition arises even when the light-matter interaction and disorder potential are, by themselves, individually Hermitian. Instead, Hermiticity arises from multiple scattering. This enables to accrue anomalous geometric phases not typically found in the simple direct electric-dipole transitions 

Such non-Hermtian shift photocurrents are particularly significant because both linear (polarized) and circular shift photocurrents have often been believed to be insensitive to scattering~\cite{WatanabeYanase2021, Kawasaki}. For example, for ultrafast  measurements it is often believed that for light frequencies $\omega \gg \tau^{-1}$, conventional (Hermitian) bulk quantum geometric photocurrents provide a good estimate of the ultrafast photocurrent measured \cite{Kawasaki,BraunEtAl2016,ReesEtAl2020}. These occur before the kinematics of relaxational processes kick in \cite{KonigEtAl2017}. Our work shows that even in this ultrafast limit, non-Hermitian disorder-assisted absorption processes dramatically affect ultrafast shift currents. It produces a qualitatively new class of non-Hermitian shift photocurrents that have opposite symmetry properties to that of the conventional shift currents. In this sense, scattering {\it enriches} photocurrent phenomena rather than diminishing it. 

At a practical level, the opposite symmetry behavior of the non-Hermitian photocurrents as compared with the conventional (symmetric) photocurrents means that the clean rule-of-thumb delineation of linear (circular) shift (injection) currents dominating the $\mathcal{T}$ ($\mathcal{PT}$) photocurrent response~\cite{WatanabeYanase2021, MaEtAl2023, AhnGuoNagaosa2020, DongEtAl2026, KangEtAl2025, XuEtAl2018} is spoiled. That said we expect a clean delineation in excitonic systems. This is because in excitonic systems, the injection and ballistic currents vanish due to the bound nature of the exciton electron-hole pair. In such systems, excitonic non-Hermitian circular shift photocurrents will dominate the circular photocurrent response in $\mathcal{T}$ excitonic systems. In contrast, excitonic Hermitian (conventional) circular shift photocurrents will dominate the circular photocurrent response in $\mathcal{PT}$. While we have focused here on the the circular non-hermitian shift photocurrent in non magnetic systems as a concrete illustration, no less important are the magnetic systems. We expect the excitonic transition non-Hermitian linear shift photocurrents will dominate the linear polarized photocurrent response of $\mathcal{T}$ excitonic systems. As such, linear photogalvanic photoresponses in excitonic PT symmetric magnets will be fruitful target to directly extract the linearly polarized effects of non-Hermitian optical transitions. 

At a deeper level, our work dramatically expands the quantum geometric description of resonant optical transitions. It is useful to note that the quantum geometry of resonant optical transitions was enumerated in Ref~\cite{AhnEtAl2022} for direct resonant transitions; these relied on the electric-dipole. By treating multiple scattering {\it indirect} transition processes, our work shows that non-Hermitian processes can possess a richer quantum geometry with both symmetric and skew resonant quantum geometry. Perhaps most striking is the fact that the linear polarized Ballistic current~\cite{BelinicherSturman1980, LeppenenGolub2023, SturmanFridkin1992, Sturman2020, Ivchenko, DaiEtAl2021, DaiRappe2021} can be directly tracked by the real part of the skew quantum metric. The linear polarized Ballistic current has consistently fallen outside of the quantum geometric paradigm and was regarded as an extrinsic contribution to photocurrents; quantum geometric photocurrents (conventional injection and shift currents) were regarded as intrinsic because they arose from bare bandstructure/wavefunction properties. Our work unifies both extrinsic scattering~\cite{BelinicherSturman1980, LeppenenGolub2023, SturmanFridkin1992, Sturman2020, Ivchenko, DaiEtAl2021, DaiRappe2021} and intrinsic quantum geometric~\cite{AhnGuoNagaosa2020, WatanabeYanase2021, AhnEtAl2022, MaEtAl2023} pictures of photocurrents by showing that both photocurrents fall within a larger scattering induced quantum geometry of non-Hermitian resonant optical transitions. Indeed, we find that geometric phase phenomena are not exclusive to the intrinsic conventional shift currents, but also manifest in multiple-scattering-off impurities processes. This is reminiscent of how skew and side jump processes in the anomalous Hall effect can be traced to an intrinsic Pancharatnam-Berry phase~\cite{Sinitsyn2008}. In non-Hermitian optical transitions, the non-Hermitian interference Wilson-loop $\mathcal{Z}_{\alpha \beta}$ plays an analogous role. Indeed, we anticipate that beyond the resonant quantum metric and quantum connection, indirect non-Hermitian optical transitions can similarly transform the geometric curvature that have been linked to photoconductivity~\cite{AhnEtAl2022, Fregoso2019, YangInteractions2026}.

Here we have focused on interband optical transitions in insulators. Interestingly, we note that for clean systems, the direct electric dipole optical transition also builds the intraband quantum geometry in metals through a ``sum over states'' formulation~\cite{qianniurmp,WatanabeYanase2021}. These sum over states formulations have proved useful in establishing strong sum rules for optical response~\cite{OnishiFu2024,WatanabeOshikawa2020,MatsyshynSodemann}. We anticipate that in much the same way, non-Hermitian optical transitions may similarly contribute to an intraband non-Hermitian quantum geometry. Particularly exciting is how these may connect with extrinsic intraband nonlinearities in the presence of a Fermi surface~\cite{Nandy2019,GolubSpivak2020,konig,isobe,zzdu1,zzdu2} as well as other multiple-scattering-induced phenomena near a Fermi surface~\cite{Sinitsyn2008}; similarly, we anticipate that virtual (disorder-assisted) non-Hermitian optical transitions may enrich intrinsic Fermi surface photocurrents~\cite{GaoRappe2021,WatanabeYanase2021} originally investigated only within the simple electric dipole approximation. Investigating these are a promising direction for future exploration of non-Hermitian transitions in both the interband setting described here and those near the Fermi surface.

\vspace{3mm}
\textit{Acknowledgements}. We thank Oles Matsyshyn for useful conversations. This research is supported by the Singapore Ministry of Education (MOE) under its Academic Research Fund Tier 2 grant MOE-T2EP50225-0008 and Academic Research Fund Tier 3 grant MOE-MOET32023-0003 ``Quantum Geometric Advantage''.

\newpage

\section*{Supplementary Material for ``Non-Hermitian Photocurrents and Skew Quantum Geometry''}
\subsection{Generalized $T$-matrix for Optical Transitions}
In this section, we derive the generalized optical $T$-matrix from the Dyson expansion and obtain the corresponding transition rates from the time evolution of transition probabilities. Comparing forward and reverse scattering paths then reveals how Wilson-loop phases combine with propagator phases to produce nonreciprocal transition rates.
\subsubsection{From Liouville evolution to the $T$-matrix}
Let $H_0\ket n=\epsilon_n\ket n$, and let each perturbation $\hat O$ transfer energy $\omega_O$. Introducing a common adiabatic switch-on parameter $\eta$, we write
\begin{equation}
 \hat{\mathbf V}^\eta(t)
 =e^{\eta t/\hbar}
 \sum_{\hat O} e^{-i\omega_{\hat O}t/\hbar}\hat O,
 \qquad \eta>0.
 \label{eq:supp_s1_V_eta}
\end{equation}
Here $\eta$ has units of energy and ensures that the perturbation vanishes as $t\to-\infty$. It serves as an adiabatic regulator, however, in the generalized situation it could also be viewed as a physical relaxation rate.

The corresponding interaction-picture perturbation is 
\begin{equation}
 \hat{\mathbf V}_I^\eta(t)
 =e^{iH_0t/\hbar}\hat{\mathbf V}^\eta(t)
 e^{-iH_0t/\hbar}.
 \label{eq:supp_s1_VI_def}
\end{equation}
Its matrix elements contain the energy mismatch between the unperturbed states and the energy transferred by the perturbation.

The interaction-picture evolution operator is
\begin{equation}
 U_I(t,-\infty)
 =\mathcal T_t\exp\!\left[
 -\frac{i}{\hbar}\int_{-\infty}^t dt'\,
 \hat{\mathbf V}_I^\eta(t')
 \right],
 \label{eq:supp_s1_UI_def}
\end{equation}
where $\mathcal T_t$ places later-time operators to the left. Expanding the exponential gives $U_I=1+\sum_{N\geq1}U_I^{(N)}$, with $U_I^{(N)}=\frac{1}{N!}(-i/\hbar)^N\mathcal T_t[(\int_{-\infty}^t dt'\,\hat{\mathbf V}_I^\eta(t'))^N]$. The integration domain contains $N!$ equivalent time-ordering sectors. Relabeling the integration variables maps each sector onto $t_1<\cdots<t_N<t$, canceling the factor $1/N!$ and yielding
\begin{equation}
\begin{aligned}[b]
 U_I^{(N)}(t,-\infty)
 ={}&\left(-\frac{i}{\hbar}\right)^N
 \int_{-\infty}^{t}dt_N
 \int_{-\infty}^{t_N}dt_{N-1}
 \cdots
 \int_{-\infty}^{t_2}dt_1\\
 &\times
 \hat{\mathbf V}_I^\eta(t_N)\cdots
 \hat{\mathbf V}_I^\eta(t_1).
\end{aligned}
\label{eq:supp_s1_UI_N}
\end{equation}
Each term describes perturbations acting from right to left, with the earliest event at $t_1$. Selecting a perturbation component at each event specifies an ordered scattering path $P=(\hat O_N,\ldots,\hat O_1)$.

For a fixed path, the time integrals are evaluated
successively using
\begin{equation}
 -\frac{i}{\hbar}\int_{-\infty}^{T}dt\,
 e^{(i\Delta+s\eta)t/\hbar}
 =
 \frac{e^{(i\Delta+s\eta)T/\hbar}}
 {-\Delta+is\eta},
 \qquad s>0.
 \label{eq:supp_s1_elementary_integral}
\end{equation}
Here $\Delta$ is the accumulated energy. The positive regulator ensures convergence at $\eta\to0$ limit and generates the retarded prescription in the energy denominators. On the other hand, $s$ counts the perturbation scatterings accumulated at that integration step.

For a transition from $\ket i$ to $\ket f$ with total transferred energy $\omega$, we have $x_\omega=\epsilon_f-\epsilon_i-\omega.$ Thus $\delta (x_\omega)$ specifies the energy-conserving final shell, with $\omega$ equal to the sum of the energy transfers along the path.

The final integration contributes the common factor
\begin{equation}
 F_\omega^{(a\eta)}(t)
 =
 \frac{e^{(ix_\omega+a\eta)t/\hbar}}
 {-x_\omega+ia\eta},
 \qquad a>0.
 \label{eq:supp_s1_F_eta}
\end{equation}
For an $N$th-order path with the common adiabatic switch-on, $a=N$. This factor isolates the final-state energy mismatch, while the remaining denominators describe propagation through intermediate states.

After the first $s$ perturbation along path $P$, define the accumulated energy transfer as $\omega_{P,s}=\sum_{j=1}^{s}\omega_{\hat O_j}$. Successive time integrations generate the retarded propagator
\begin{equation}
 G_{P,s}^{\rm ret,\eta}
 =(\epsilon_i+\omega_{P,s}-H_0+is\eta)^{-1}.
 \label{eq:supp_s1_path_propagator}
\end{equation}
The propagator energy includes all preceding energy transfers, while $s\eta$ accumulates the adiabatic regulators from the first $s$ scatterings. At the same time, for a path containing $N$ scatterings, the transition amplitude could be factorized as
\begin{equation}
 c_{P;fi}^{(\eta)}(t)
 =F_{\omega_{P,N}}^{(N\eta)}(t)
 \mathcal A_{P;fi}^{(\eta)}.
 \label{eq:supp_s1_path_factorization}
\end{equation}
The final-shell factor contains the total energy transfer and the accumulated regulator $N\eta$, which separates the explicit time dependence from the reduced path amplitude, where the reduced amplitude is
\begin{equation}
 \mathcal A_{P;fi}^{(\eta)}
 =\langle f|
 \hat O_NG_{P,N-1}^{\rm ret,\eta}\hat O_{N-1}
 \cdots G_{P,1}^{\rm ret,\eta}\hat O_1
 |i\rangle.
 \label{eq:supp_s1_ordered_factorization}
\end{equation}
Each intermediate propagator connects two successive perturbations in order. Taking its Hermitian adjoint gives the advanced propagator, $G_{P,s}^{A,\eta}=(G_{P,s}^{\rm ret,\eta})^\dagger$.

Therefore, the $T$-matrix is obtained by summing the reduced path amplitudes with fixed initial and final states and total energy transfer. Define $G_\eta^{\rm ret}(E)=(E-H_0+i\eta)^{-1}$ where $\eta\to0^+$, and $\mathcal A_{P;fi}=\lim_{\eta\to0^+}\mathcal A_{P;fi}^{(\eta)}$. Writing $\omega_P=\omega_{P,N}$, we obtain
\begin{equation}
 T_{\omega;fi}
 =\langle f|\hat T_\omega(\epsilon_i)|i\rangle
 =\sum_{P:\omega_P=\omega}\mathcal A_{P;fi}.
 \label{eq:supp_s1_process_T}
\end{equation}
All paths with the same net energy transfer contribute simultaneously to this transition amplitude.

For a fixed initial energy $E=\epsilon_i$, the corresponding operator yields the general $T$-matrix expansion given in the main text.
\begin{equation}
\begin{aligned}[b]
 \hat T_\omega(E)
 ={}&\sum_{N\ge1}
 \sum_{\substack{\hat O_1,\ldots,\hat O_N\\
                 \omega_{P,N}=\omega}}
 \hat O_NG_\eta^{\rm ret}(E+\omega_{P,N-1})
 \\
 &\qquad\times\hat O_{N-1}
 \cdots G_\eta^{\rm ret}(E+\omega_{P,1})\hat O_1 .
\end{aligned}
\label{eq:supp_s1_process_T_operator}
\end{equation}
Each denominator records the energy available after the preceding scatterings. The constraint $\omega_{P,N}=\omega$ selects the desired energy-transfer process without restricting the intermediate scattering sequence.

Particularly, for a static perturbation $\hat U$, all energy transfers vanish, and the expansion reduces to the usual Born series, equivalently $\hat T(E)=\hat U+\hat UG_\eta^{\rm ret}(E)\hat T(E)$. All propagators have the same energy argument $E$, with the retarded prescription imposing the outgoing-wave boundary condition.

On the other hand, for a static perturbation $\hat U$ and an optical perturbation $\hat V_\omega^a$ transferring energy $\omega$, retaining exactly one optical perturbation which just gives us the field-free optical $T$-matrix in the main text in form of
\begin{equation}
\begin{aligned}[b]
 \hat t_\omega^a(E)
 =\sum_{m,n\ge0}
 \bigl[\hat UG_\eta^{\rm ret}(E+\omega)\bigr]^m
 \hat V_\omega^a
 \bigl[G_\eta^{\rm ret}(E)\hat U\bigr]^n.
\end{aligned}
\label{eq:supp_optical_T_series}
\end{equation}
Here $n$ static scatterings precede the optical transition and $m$ follow it, so the propagator energy changes from $E$ to $E+\omega$ at the optical scattering perturbation. The double sum includes all placements of this scattering perturbation: $m=n=0$ gives the clean optical transition, while the remaining terms describe disorder-assisted processes whose interference determines the transition rates.

\subsubsection{From transition probability to scattering rate $W$}
For $f\ne i$, the transition amplitude is $c_{fi}^{(\eta)}(t)=\sum_P\mathcal A_{P;fi}^{(\eta)}F_{\omega_P}^{(N_P\eta)}(t)$. Here $N_P$ is the number of perturbation scatterings along path $P$, and $\omega_P$ is the total energy transferred. The amplitude coherently sums all paths connecting the same initial and final states. The corresponding transition probability is
\begin{equation}
 P_{f\leftarrow i}^{(\eta)}(t)
 =|c_{fi}^{(\eta)}(t)|^2.
 \label{eq:supp_s1_probability_two_branches}
\end{equation}
Squaring the amplitude introduces interference between path $P$ and the conjugate path $Q$, with their respective adiabatic regulators retained.

The transition rate is extracted from time-averaged, part of this probability:
\begin{equation}
 W_{f\leftarrow i}
 =\lim_{\eta\to0^+}
 \partial_tP_{f\leftarrow i}^{(\eta)}(t).
 \label{eq:supp_s1_rate_as_probability_derivative}
\end{equation}
The timer averaged part retains the slowly accumulating contribution after oscillations between distinct energy-transfer processs have been averaged out. Here in transition scattering, the energy-conserving limit is understood distributionally. Specifically, pairs with $\omega_P\ne\omega_Q$ contain the oscillatory factor $e^{i(\omega_Q-\omega_P)t/\hbar}$. Averaging over times long compared with their beat periods but short compared with the adiabatic switching time removes these terms. Thus,
\begin{equation}
\begin{aligned}[b]
 P_{f\leftarrow i}^{(\eta)}(t)
 =\sum_{\pm \omega}
 \sum_{\substack{P,Q\\\omega_P=\omega_Q=\omega}}
 (\mathcal A_{Q;fi}^{(\eta)})^*
 \mathcal A_{P;fi}^{(\eta)}
 (F_\omega^{(N_Q\eta)})^*
 F_\omega^{(N_P\eta)}.
\end{aligned}
\label{eq:supp_secular_probability}
\end{equation}
However, paths with the same total energy transfer still interfere, even when they contain different numbers or orderings of perturbation scatterings.

For an equal-transfer pair, let $a=N_Q$, $b=N_P$, and $x=\epsilon_f-\epsilon_i-\omega$. The oscillatory phases cancel, giving
\begin{equation}
 (F_\omega^{(a\eta)}(t))^*
 F_\omega^{(b\eta)}(t)
 =
 \frac{e^{(a+b)\eta t/\hbar}}
 {(x+ia\eta)(x-ib\eta)}.
 \label{eq:supp_s1_F_product}
\end{equation}
The two denominators retain the advanced and retarded prescriptions, while the remaining time dependence comes entirely from the adiabatic factors.

Differentiation therefore yields
\begin{equation}
\begin{aligned}[b]
 \partial_t\!\left[
 (F_\omega^{(a\eta)}(t))^*
 F_\omega^{(b\eta)}(t)
 \right]
 =
 \frac{(a+b)\eta\,e^{(a+b)\eta t/\hbar}}
 {\hbar(x+ia\eta)(x-ib\eta)}.
\end{aligned}
\label{eq:supp_s1_F_product_derivative}
\end{equation}
For fixed observation time $t$, the exponential tends to unity as $\eta\to0^+$, while the remaining factor concentrates at the energy-conserving condition $x=0$. Which means for every $a,b>0$, the distributional limit is
\begin{equation}
 \lim_{\eta\to0^+}
 \partial_t\!\left[
 (F_\omega^{(a\eta)}(t))^*
 F_\omega^{(b\eta)}(t)
 \right]
 =
 \frac{2\pi}{\hbar}\delta(x).
 \label{eq:supp_s1_eta_delta_general_ab}
\end{equation}
This follows by partial fractions and the identity $(x\pm i0^+)^{-1}=\mathcal P(1/x)\mp i\pi\delta(x)$. The resulting energy-conserving prefactor is independent of the two path orders; their interference remains encoded in $(\mathcal A_{Q;fi}^{(\eta)})^*\mathcal A_{P;fi}^{(\eta)}$.

Expanding the scalar transition rate in the field bilinears $E_\omega^aE_{-\omega}^b$ gives the field-independent rate tensor
\begin{equation}
 W_{\omega;\beta\alpha}^{ab}
 =\frac{2\pi e^2}{\hbar}
 \delta(\epsilon_\beta-\epsilon_\alpha-\omega)
 t_{\omega;\beta\alpha}^a
 (t_{\omega;\beta\alpha}^b)^*.
 \label{eq:supp_s1_optical_rate_fieldfree}
\end{equation}
Contracted with $E_\omega^aE_{-\omega}^b$, we will obtain Eq.~\eqref{eq:rate} in the main text. For a real electric field, $E_{-\omega}^b=(E_\omega^b)^*$. Thus, $-\omega$ labels the conjugate field component, while the conjugate field-independent amplitude is $(t_{\omega;\beta\alpha}^b)^*$, associated with the same transition process.

\subsubsection{Generalized Wilson-loop Obstruction}
We set $\hbar=1$ and let $\hat H_0\ket{i}=\epsilon_i\ket{i}$. A general Hermitian perturbation can be written as
\begin{equation}
 \hat{\mathbf V}(t)
 =\hat O_0+\sum_{\omega>0}
 \left(
 \hat O_\omega e^{-i\omega t}
 +\hat O_\omega^\dagger e^{i\omega t}
 \right).
 \label{eq:herm}
\end{equation}
Here $\hat O_0=\hat O_0^\dagger$, while $\hat O_\omega$ and $\hat O_\omega^\dagger$ transfer energies $+\omega$ and $-\omega$, respectively. The retarded $T$-matrix can then be expanded, as in the main text, in form of
\begin{equation}
\begin{aligned}[b]
 \hat T
 =\hat{\mathbf V}
 +\hat{\mathbf V}G_\eta^{\rm ret}\hat{\mathbf V}
 +\hat{\mathbf V}G_\eta^{\rm ret}\hat{\mathbf V}
 G_\eta^{\rm ret}\hat{\mathbf V}+\cdots .
\end{aligned}
\label{eq:T}
\end{equation}
Here $G_\eta^{\rm ret}(E)=(E-\hat H_0+i\eta)^{-1}$, with the energy determined by the preceding energy transfers along each path.

For a transition $\alpha\to\beta$ with net energy transfer $\omega$, the frequency-resolved transition rate is
\begin{equation}
 W_{\alpha\beta}^{(\omega)}
 \equiv W_{\alpha\to\beta}^{(\omega)}
 =2\pi\Delta_\omega
 \left|T_{\beta\alpha}(\omega)\right|^2.
 \label{eq:supp-wilson-rate}
\end{equation}
The factor $\Delta_\omega=\delta(\epsilon_\beta-\epsilon_\alpha-\omega)$ enforces overall energy conservation. The total rate is $W_{\alpha\beta}=\sum_\omega W_{\alpha\beta}^{(\omega)}$, including both absorption and emission processes. Restoring $\hbar$ recovers Eq.~\eqref{eq:rate} in the main text. Since the reverse process transfers energy $-\omega$, the antisymmetric and symmetric frequency-resolved rates are defined as $W_{\alpha\beta}^{A,(\omega)}=[W_{\alpha\beta}^{(\omega)}-W_{\beta\alpha}^{(-\omega)}]/2$ and $W_{\alpha\beta}^{S,(\omega)}=[W_{\alpha\beta}^{(\omega)}+W_{\beta\alpha}^{(-\omega)}]/2$, respectively. These definitions underlie the derivation of Eq.~\eqref{eq:antirate}.

An $r$-perturbation contribution to $T_{\beta\alpha}(\omega)$ follows an ordered path $P=(i_0=\alpha;\hat O_1,i_1;\ldots;\hat O_r,i_r=\beta)$. Each perturbation transfers energy $\nu_j$ ($+\omega_j$ for $\hat O_{\omega_j}$, $-\omega_j$ for $\hat O_{\omega_j}^\dagger$, and zero for $\hat O_0$), with $\sum_{j=1}^r\nu_j=\omega$. Define the numerator product $\boldsymbol{\mathcal M}_P=\prod_{j=1}^r\langle i_j|\hat O_j|i_{j-1}\rangle$ and the retarded propagator product $\boldsymbol{\mathcal C}_P^{\rm ret}=\prod_{m=1}^{r-1}(E_m-\epsilon_{i_m}+i\eta)^{-1}$, where $E_m=\epsilon_\alpha+\sum_{j=1}^m\nu_j$. The intermediate states need not be on shell, and the empty propagator product equals one. Summing over all path lengths and orderings gives
\begin{equation}
 T_{\beta\alpha}(\omega)
 =\sum_{r\ge1}\sum_{P_r:\omega}
 \boldsymbol{\mathcal C}_{P_r}^{\rm ret}
 \boldsymbol{\mathcal M}_{P_r}.
 \label{eq:allorder_amplitude}
\end{equation}
which gives out Eq.~\eqref{transition rate pathes expansion}Paths with the same net energy transfer contribute coherently, including interference between different perturbation orders. Their relative phases involve both the numerator matrix elements and the retarded propagators.

Hence, the corresponding rate is
\begin{equation}
 W_{\alpha\beta}^{(\omega)}
 =2\pi\Delta_\omega
 \sum_{r,s\ge1}\sum_{P_r,Q_s:\omega}
 \boldsymbol{\mathcal C}_{P_r}^{\rm ret}
 \boldsymbol{\mathcal C}_{Q_s}^{\rm adv}
 \boldsymbol{\mathcal M}_{P_r}
 \boldsymbol{\mathcal M}_{Q_s}^*.
\label{eq:allorder}
\end{equation}
Here $\boldsymbol{\mathcal C}_Q^{\rm adv}=(\boldsymbol{\mathcal C}_Q^{\rm ret})^*$. The sum includes all ordered path pairs, including interference between different scattering orders, so no additional factor of $1/2$ is introduced. According to this equation, we could subsequently derive Eq.~\eqref{eq:injection-rate-u2}

Reversing the perturbation sequence and taking each perturbation's Hermitian adjoint gives $P^\dagger=(i_r=\beta;\hat O_r^\dagger,i_{r-1};\ldots;\hat O_1^\dagger,i_0=\alpha)$. This path transfers energy $-\omega$ and satisfies $\boldsymbol{\mathcal M}_{P^\dagger}=\boldsymbol{\mathcal M}_P^*$. The reverse transition is still evaluated with retarded propagators.

On the energy shell $\epsilon_\beta=\epsilon_\alpha+\omega$, the energy available when the reverse path reaches $i_m$ is
\begin{equation}
 E_{P^\dagger,r-m}
 =\epsilon_\beta-\sum_{j=m+1}^r\nu_j
 =\epsilon_\alpha+\sum_{j=1}^m\nu_j
 =E_m.
 \label{eq:cut}
\end{equation}
Thus, the forward and reverse paths have identical denominator products, $\boldsymbol{\mathcal C}_{P^\dagger}^{\rm ret}=\boldsymbol{\mathcal C}_P^{\rm ret}$ and $\boldsymbol{\mathcal C}_{P^\dagger}^{\rm adv}=\boldsymbol{\mathcal C}_P^{\rm adv}$, giving out Eq.~\eqref{eq:reverse-path-propagators}. Meanwhile their perturbation products are complex conjugates. Their energy-conserving delta functions also coincide, since $\delta(\epsilon_\alpha-\epsilon_\beta+\omega)=\Delta_\omega$.

Define $\boldsymbol{\mathcal L}_{P,Q}=\boldsymbol{\mathcal M}_P\boldsymbol{\mathcal M}_Q^*$ and $\boldsymbol{\mathcal C}_{P,Q}=\boldsymbol{\mathcal C}_P^{\rm ret}\boldsymbol{\mathcal C}_Q^{\rm adv}$. For $P\ne Q$, combining $(P,Q)$ with its conjugate $(Q,P)$ gives the forward and reverse contributions
\begin{equation}
\begin{aligned}
 \mathcal F_{P,Q}^{\alpha\to\beta}
 &=4\pi\Delta_\omega
 \operatorname{Re}\!\left(
 \boldsymbol{\mathcal L}_{P,Q}\boldsymbol{\mathcal C}_{P,Q}
 \right),\\
 \mathcal F_{P,Q}^{\beta\to\alpha}
 &=4\pi\Delta_\omega
 \operatorname{Re}\!\left(
 \boldsymbol{\mathcal L}_{P,Q}^*\boldsymbol{\mathcal C}_{P,Q}
 \right).
\end{aligned}
\label{eq:loop_forward_reverse}
\end{equation}
The two processes transfer energies $\omega$ and $-\omega$, respectively, with the same energy-conserving factor $\Delta_\omega=\delta(\epsilon_\beta-\epsilon_\alpha-\omega)$. Loop reversal conjugates the numerator product while leaving the propagator products unchanged.

Their antisymmetric and symmetric combinations are therefore
\begin{equation}
\begin{aligned}
 \frac{
 \mathcal F_{P,Q}^{\alpha\to\beta}
 -\mathcal F_{P,Q}^{\beta\to\alpha}}{2}
 &=-4\pi\Delta_\omega
 \operatorname{Im}\boldsymbol{\mathcal L}_{P,Q}
 \operatorname{Im}\boldsymbol{\mathcal C}_{P,Q},\\
 \frac{
 \mathcal F_{P,Q}^{\alpha\to\beta}
 +\mathcal F_{P,Q}^{\beta\to\alpha}}{2}
 &=4\pi\Delta_\omega
 \operatorname{Re}\boldsymbol{\mathcal L}_{P,Q}
 \operatorname{Re}\boldsymbol{\mathcal C}_{P,Q}.
\end{aligned}
\label{eq:loop_skew_sym}
\end{equation}
An antisymmetric contribution requires nonzero imaginary parts of both the numerator and propagator products, whereas the symmetric contribution involves their real parts.

Re-expressing these paired contributions as sums over all ordered path pairs, including both $(P,Q)$ and $(Q,P)$, gives
\begin{align}
 W_{\alpha\beta}^{A}(\omega)
 &=-2\pi\Delta_\omega
 \sum_{P,Q}
 \operatorname{Im}\boldsymbol{\mathcal L}_{P,Q}
 \operatorname{Im}\boldsymbol{\mathcal C}_{P,Q},
 \label{eq:result}\\
 W_{\alpha\beta}^{S}(\omega)
 &=2\pi\Delta_\omega
 \sum_{P,Q}
 \operatorname{Re}\boldsymbol{\mathcal L}_{P,Q}
 \operatorname{Re}\boldsymbol{\mathcal C}_{P,Q}.
 \label{eq:result_sym}
\end{align}
The ordered sums include the diagonal terms $P=Q$, which vanish in the anti-symmetric rate but contribute once to the symmetric rate. In particular, the symmetric rate survives in the single-perturbation limit, where $\boldsymbol{\mathcal C}_P^{\rm ret}=1$. For the anti-symmetric rate, opposite loop orientations carry conjugate perturbation products and hence opposite Wilson-loop phases \cite{Sinitsyn2008,MaAroraVignaleSong2023,ShiSong2019}. Their interference with the propagator phases can produce unequal forward and reverse rates. We refer to this obstruction to reciprocity as the generalized Wilson-loop obstruction.

We now specialize the preceding results to static Hermitian disorder $\hat U$ and an optical perturbation $\hat O_\omega=-eE^a(\omega)\hat r^a$, retaining exactly one optical perturbation in each amplitude. As above, $\alpha$ and $\beta$ denote the initial and final states, respectively. Each path numerator factors as $\boldsymbol{\mathcal M}_P=-eE^a(\omega)\boldsymbol{\mathcal M}_P^a$, where $\boldsymbol{\mathcal M}_P^a$ contains one matrix element of $\hat r^a$ and otherwise only disorder matrix elements. Consequently,
\begin{equation}
 t_{\beta\alpha}^{a}(\omega;\epsilon_\alpha)
 =\sum_{P:\omega}
 \boldsymbol{\mathcal C}_P^{\rm ret}
 \boldsymbol{\mathcal M}_P^a.
 \label{eq:supp_opt_path_amplitude}
\end{equation}
The denominator products are independent of polarization. Applying Eq.~\eqref{eq:cut} to the reverse process gives $\boldsymbol{\mathcal C}_{P^\dagger}^{\rm ret}=\boldsymbol{\mathcal C}_P^{\rm ret}$, $\boldsymbol{\mathcal C}_{P^\dagger}^{\rm adv}=\boldsymbol{\mathcal C}_P^{\rm adv}$, and $\boldsymbol{\mathcal M}_{P^\dagger}^a=(\boldsymbol{\mathcal M}_P^a)^*$. Since the reverse process also carries the conjugate field, its field-contracted loop amplitude is the complex conjugate of the forward one, establishing Eq.~\eqref{phase relation}.

Define $\boldsymbol{\mathcal L}_{P,Q}^{ab}=\boldsymbol{\mathcal M}_P^a(\boldsymbol{\mathcal M}_Q^b)^*$ and $\boldsymbol{\mathcal C}_{P,Q}=\boldsymbol{\mathcal C}_P^{\rm ret}\boldsymbol{\mathcal C}_Q^{\rm adv}$. Substituting the field-factorized numerators into Eqs.~\eqref{eq:result} and~\eqref{eq:result_sym}, and using $E^a(\omega)E^b(-\omega)=L^{ab}(\omega)+iH^{ab}(\omega)$, yields $W_{\alpha\beta}^{A}(\omega)=$
\begin{equation}
\begin{aligned}[b]
 -2\pi e^2\Delta_\omega\sum_{a,b,P,Q}
 \Big[
 L^{ab}_\omega\operatorname{Im}\boldsymbol{\mathcal L}_{P,Q}^{ab}+H^{ab}_\omega\operatorname{Re}\boldsymbol{\mathcal L}_{P,Q}^{ab}
 \Big]
 \operatorname{Im}\boldsymbol{\mathcal C}_{P,Q},
\end{aligned}
\label{eq:supp_opt_skew_rate}
\end{equation}
and $ W_{\alpha\beta}^{S}(\omega)=$
\begin{equation}
\begin{aligned}[b]
2\pi e^2\Delta_\omega\sum_{a,b,P,Q}
 \Big[
 L^{ab}_\omega\operatorname{Re}\boldsymbol{\mathcal L}_{P,Q}^{ab}-H^{ab}_\omega\operatorname{Im}\boldsymbol{\mathcal L}_{P,Q}^{ab}
 \Big]
 \operatorname{Re}\boldsymbol{\mathcal C}_{P,Q}.
\end{aligned}
\label{eq:supp_opt_symmetric_rate}
\end{equation}
Here the reverse rate entering $W^{S,A}$ belongs to the $-\omega$ process. All path-pair sums remain ordered, so no additional pairing factor is required. Restoring $\hbar$ replaces $2\pi$ by $2\pi/\hbar$, consistently with the rates in the main text.

To connect these results to $\mathcal Z_{\alpha\beta}^{ab}(\omega)$, abbreviate the reciprocal amplitudes as $t_{\beta\alpha}^a=t_{\beta\alpha}^{a,\rm ret}(\omega;\epsilon_\alpha)$ and $t_{\alpha\beta}^a=t_{\alpha\beta}^{a,\rm ret}(-\omega;\epsilon_\beta)$. The process-paired Hermitian decomposition gives $\mathcal X_{\alpha\beta}^{b,-\omega}=[t_{\alpha\beta}^b+(t_{\beta\alpha}^b)^*]/2=\sum_Q(\boldsymbol{\mathcal M}_Q^b)^*\operatorname{Re}\boldsymbol{\mathcal C}_Q^{\rm ret}$ and $\mathcal Y_{\beta\alpha}^{a,\omega}=[t_{\beta\alpha}^a-(t_{\alpha\beta}^a)^*]/(2i)=\sum_P\boldsymbol{\mathcal M}_P^a\operatorname{Im}\boldsymbol{\mathcal C}_P^{\rm ret}$. Hence
\begin{equation}
 \mathcal Z_{\alpha\beta}^{ab}(\omega)
 =\sum_{P,Q}\boldsymbol{\mathcal L}_{P,Q}^{ab}
 \operatorname{Im}\boldsymbol{\mathcal C}_P^{\rm ret}
 \operatorname{Re}\boldsymbol{\mathcal C}_Q^{\rm ret}.
 \label{eq:supp_opt_Z_paths}
\end{equation}
Using $\boldsymbol{\mathcal L}_{Q,P}^{ba}=(\boldsymbol{\mathcal L}_{P,Q}^{ab})^*$, relabeling $P\leftrightarrow Q$, and separating the symmetric and antisymmetric polarization components gives
\begin{align}
 \operatorname{Im}\mathcal Z_{\alpha\beta}^{(ab)}(\omega)
 &=\frac12\sum_{P,Q}
 \operatorname{Im}\boldsymbol{\mathcal L}_{P,Q}^{(ab)}
 \operatorname{Im}\boldsymbol{\mathcal C}_{P,Q},
 \label{eq:supp_opt_Z_im_linear}\\
 \operatorname{Re}\mathcal Z_{\alpha\beta}^{[ab]}(\omega)
 &=\frac12\sum_{P,Q}
 \operatorname{Re}\boldsymbol{\mathcal L}_{P,Q}^{[ab]}
 \operatorname{Im}\boldsymbol{\mathcal C}_{P,Q},
 \label{eq:supp_opt_Z_re_circular}\\
 \operatorname{Re}\mathcal Z_{\alpha\beta}^{(ab)}(\omega)
 &=\frac12\sum_{P,Q}
 \operatorname{Im}\!\left[
 \boldsymbol{\mathcal L}_{P,Q}^{(ab)}
 \boldsymbol{\mathcal C}_P^{\rm ret}
 \boldsymbol{\mathcal C}_Q^{\rm ret}
 \right],
 \label{eq:supp_opt_Z_re_linear}\\
 -\operatorname{Im}\mathcal Z_{\alpha\beta}^{[ab]}(\omega)
 &=\frac12\sum_{P,Q}
 \operatorname{Re}\!\left[
 \boldsymbol{\mathcal L}_{P,Q}^{[ab]}
 \boldsymbol{\mathcal C}_P^{\rm ret}
 \boldsymbol{\mathcal C}_Q^{\rm ret}
 \right].
 \label{eq:supp_opt_Z_im_circular}
\end{align}
The factors of $1/2$ follow from averaging the expressions before and after exchanging the ordered paths. Combining the symmetric and antisymmetric polarization components yields
\begin{align}
 \operatorname{Im}\mathcal Z_{\alpha\beta}^{ab,\omega}
 &\hspace{-1mm}=\frac12\sum_{P,Q}\Big\{
 \operatorname{Im}\boldsymbol{\mathcal L}_{P,Q}^{(ab)}
 \operatorname{Im}\boldsymbol{\mathcal C}_{P,Q}
 \hspace{-1mm}-\hspace{-1mm}\operatorname{Re}\!\left[
 \boldsymbol{\mathcal L}_{P,Q}^{[ab]}
 \boldsymbol{\mathcal C}_P^{\rm ret}
 \boldsymbol{\mathcal C}_Q^{\rm ret}
 \right]\Big\},
 \label{eq:supp_opt_Z_im_full}\\
 \operatorname{Re}\mathcal Z_{\alpha\beta}^{ab,\omega}
 &\hspace{-1mm}=\frac12\sum_{P,Q}\Big\{
 \operatorname{Re}\boldsymbol{\mathcal L}_{P,Q}^{[ab]}
 \operatorname{Im}\boldsymbol{\mathcal C}_{P,Q}
 \hspace{-1mm}+\hspace{-1mm}\operatorname{Im}\!\left[
 \boldsymbol{\mathcal L}_{P,Q}^{(ab)}
 \boldsymbol{\mathcal C}_P^{\rm ret}
 \boldsymbol{\mathcal C}_Q^{\rm ret}
 \right]\Big\}.
 \label{eq:supp_opt_Z_re_full}
\end{align}
These expressions reproduce Eqs.~\eqref{anti identity} and~\eqref{eq:anti-identity-circular} in the main text. Contracting with $L^{ab}(\omega)$ and $H^{ab}(\omega)$ then recovers the antisymmetric transition rate. Likewise, the same expressions for $\mathcal X$ and $\mathcal Y$ give
\begin{equation}
\begin{aligned}[b]
 \mathcal X_{\alpha\beta}^{b,-\omega}
 \mathcal X_{\beta\alpha}^{a,\omega}
 +\mathcal Y_{\alpha\beta}^{b,-\omega}
 \mathcal Y_{\beta\alpha}^{a,\omega}=\sum_{P,Q}
 \boldsymbol{\mathcal L}_{P,Q}^{ab}
 \operatorname{Re}\boldsymbol{\mathcal C}_{P,Q}.
\end{aligned}
\label{eq:supp_opt_symmetric_kernel}
\end{equation}
Taking its real symmetric and imaginary antisymmetric polarization components reproduces the two symmetric-kernel identities in the main text and, upon field contraction, Eq.~\eqref{eq:supp_opt_symmetric_rate}. In the clean limit, $\boldsymbol{\mathcal C}_0^{\rm ret}=1$ and $\boldsymbol{\mathcal M}_0^a=\langle\beta|\hat r^a|\alpha\rangle$, so $\mathcal Y^a=\mathcal Z^{ab}=0$, while the symmetric rate survives. This completes the specialization of the general proof to disorder-assisted optical transitions.

\subsection{Generalized $T$-matrix Shift Vector} 
In this section we derive the generalized T-matrix shift vector from a wavepacket analysis at fixed momentum transfer, allowing different initial and final momenta. We then determine the $\mathcal T$ and $\mathcal{PT}$ transformation properties of its symmetric and antisymmetric components for linear and circular polarizations.

\subsubsection{Derivation of Generalized $T$-matrix Shift Vector}
In what follows, we will use Bloch states with $|\psi_{a\mathbf k}\rangle=e^{i\mathbf k\cdot\hat{\mathbf r}}|u_a(\mathbf k)\rangle$ and Berry connection $\boldsymbol{\mathcal A}_a(\mathbf k)=i\langle u_a(\mathbf k)|
\nabla_{\mathbf k}u_a(\mathbf k)\rangle$. For a specified scattering or optical process, define $T_{mn}(\mathbf k_f,\mathbf k_i)=\langle\psi_{m\mathbf k_f}|\widehat{T}|\psi_{n\mathbf k_i}\rangle=|T_{mn}(\mathbf k_f,\mathbf k_i)|e^{i\phi_{mn}(\mathbf k_f,\mathbf k_i)}.$ We will first examine the incident wavepacket as $|\Psi_i\rangle=\int_{\mathbf q_i}f_i(\mathbf q_i)|\psi_{n\mathbf q_i}\rangle$, where $\int_{\mathbf q}\equiv\int d^dq/(2\pi)^d$. We choose $f_i$ to be real, nonnegative, and sharply peaked at $\mathbf k_i$. Following Ref.~\cite{Xiong2022Thesis} (see also Refs.~\cite{ShiSong2019,ShiPlasmon2018} ), its probability density is
\begin{equation}
\begin{aligned}[b]
 |\Psi_i(\mathbf r)|^2=
 \int_{\mathbf q_i,\mathbf p}
 f_i(\mathbf q_i)f_i(\mathbf q_i+\mathbf p)
 \langle u_n(\mathbf q_i)|
 u_n(\mathbf q_i+\mathbf p)\rangle
 e^{i\mathbf p\cdot\mathbf r}.
\end{aligned}
\label{eq:sv-indensity}
\end{equation}
The overlap accounts for the momentum dependence of
the Bloch basis, while the real envelope contributes
no additional phase. Utilizing $\langle u_a(\mathbf q)|u_a(\mathbf q+\mathbf p)\rangle=\exp[-i\mathbf p\cdot\boldsymbol{\mathcal A}_a(\mathbf q)+O(p^2)]$, we obtain a phase factor
\begin{equation}
 \chi_i
 =\mathbf p\cdot
 [\mathbf r-\boldsymbol{\mathcal A}_n(\mathbf q_i)]
 +O(p^2).
 \label{eq:sv-inphase}
\end{equation}
This means that the initial position will depend on the Berry connection. To see this we apply the stationary-phase method: 
$\left.\nabla_{\mathbf p}\chi_i\right|_{\mathbf p=0,\,\mathbf q_i=\mathbf k_i,\,\mathbf r=\mathbf r_i}=0$ which gives
\begin{equation}
 \mathbf r_i
 =\boldsymbol{\mathcal A}_n(\mathbf k_i),
 \label{eq:sv-incenter}
\end{equation}
that determines the center of the initial wavepacket. We note that this initial wavepacket position is not gauge invariant, but as we will see, the difference in initial and final wavepacket positions is gauge invariant. .

We next focus on the wavepacket in the final state (after the indirect optical transition) for a fixed momentum transfer $\mathbf K=\mathbf k_f-\mathbf k_i$. Writing $\mathbf q_f=\mathbf q_i+\mathbf K$, the final wavepacket is 
\begin{equation}
 |\Psi_f\rangle
 =\int_{\mathbf q_i}
 T_{mn}(\mathbf q_i+\mathbf K,\mathbf q_i)
 f_i(\mathbf q_i)
 |\psi_{m,\mathbf q_i+\mathbf K}\rangle, 
 \label{eq:sv-outpacket}
\end{equation}
where each momentum in the initial wavepacket is shifted by $\mathbf K$ and weighted by the corresponding transition amplitude. For a narrow wavepacket, retaining the interference phase through first order in $\mathbf p$ gives
\begin{equation}
\begin{aligned}[b]
  |\Psi_i(\mathbf r)|^2 = \int_{\mathbf q_i,\mathbf p}
 &f_i(\mathbf q_i)f_i(\mathbf q_i+\mathbf p)|T_{mn}(\mathbf q_i+\mathbf K,\mathbf q_i)|
 \\
 &\times
 |T_{mn}(\mathbf q_i+\mathbf K+\mathbf p,
         \mathbf q_i+\mathbf p)|
 e^{i\chi_f}, 
\end{aligned}
\label{eq:sv-outdensity}
\end{equation}
where the transition amplitudes modify both the envelope and its phase. Evaluating the phase near the peak yields
\begin{equation}
\begin{aligned}[b]
 \chi_f={}&
 \mathbf p\cdot
 [\mathbf r-\boldsymbol{\mathcal A}_m(\mathbf k_f)]
 +\phi_{mn}(\mathbf k_f+\mathbf p,\mathbf k_i+\mathbf p)
 \\
 &-\phi_{mn}(\mathbf k_f,\mathbf k_i)
 +O(p^2), 
\end{aligned}
\label{eq:sv-outphase}
\end{equation}
were the Berry connection originates from $\langle u_m(\mathbf k_f)|u_m(\mathbf k_f+\mathbf p)\rangle$. At fixed $\mathbf K$, both momentum arguments vary together, so the transition-phase variation is governed by $\nabla_{\mathbf k_f}+\nabla_{\mathbf k_i}$. 

To see this explicitly, we note $\delta \phi_{\rm mn} = \phi_{mn}(\mathbf k_f+\mathbf p,\mathbf k_i+\mathbf p)
 -\phi_{mn}(\mathbf k_f,\mathbf k_i)$ can be written in a more suggestive fashion by adding and subtracting the 
intermediate phase $\phi_{mn}(\mathbf k_f,\mathbf k_i+\mathbf p)$.  
The total phase change then satisfies the identity
\begin{equation}
\begin{aligned}[b]
 \delta\phi_{mn}
 ={}&
 \big[
 \phi_{mn}(\mathbf k_f+\mathbf p,\mathbf k_i+\mathbf p)
 -\phi_{mn}(\mathbf k_f,\mathbf k_i+\mathbf p)
 \big]
 \\
 &+
 \big[
 \phi_{mn}(\mathbf k_f,\mathbf k_i+\mathbf p)
 -\phi_{mn}(\mathbf k_f,\mathbf k_i)
 \big].
\end{aligned}
\label{eq:sv-split}
\end{equation}
The first bracket varies only the final momentum, while the second varies only the initial momentum. By expanding both brackets to first order, we obtain
\begin{equation}
 \delta\phi_{mn}
 =\mathbf p\cdot
 \bigl(\nabla_{\mathbf k_f}+\nabla_{\mathbf k_i}\bigr)
 \phi_{mn}
 +O(p^2).
 \label{eq:sv-expand}
\end{equation}
Both derivatives are evaluated at $(\mathbf k_f,\mathbf k_i)$ and act on their respective arguments. Their sum follows from varying both momenta at fixed transfer $\mathbf K$. Finally substituting Eq.~\eqref{eq:sv-expand} into Eq.~\eqref{eq:sv-outphase} yields
\begin{equation}
\begin{aligned}[b]
 \chi_f
 =\mathbf p\cdot\Big[
 \mathbf r-\boldsymbol{\mathcal A}_m(\mathbf k_f)+
 \bigl(\nabla_{\mathbf k_f}+\nabla_{\mathbf k_i}\bigr)
 \phi_{mn}
 \Big]+O(p^2).
\end{aligned}
\label{eq:sv-phase-linear}
\end{equation}
The outgoing interference phase therefore combines the final-state Berry connection with the phase variation of the transition amplitude.

Similar to how the initial wavepacket position was evaluated, we find the outgoing packet center follows from the stationary-phase method $\left.\nabla_{\mathbf p}\chi_f \right|_{\mathbf p=0,\,\mathbf r=\mathbf r_f}=0$, giving
\begin{equation}
 \mathbf r_f
 =\boldsymbol{\mathcal A}_m(\mathbf k_f)
 -\bigl(\nabla_{\mathbf k_f}+\nabla_{\mathbf k_i}\bigr)
 \phi_{mn},
 \label{eq:sv-endpoints}
\end{equation}
determining the final wavepacket center to leading order in the packet width, using the same real-envelope convention as for the incident packet.

Finally, subtracting the initial position given in Eq.~\eqref{eq:sv-incenter} yields the coordinate shift vector $\mathbf R_{mn}=\mathbf r_f-\mathbf r_i$, as given in Eq.~\eqref{eq:shift} of the main text:
\begin{equation}
\begin{aligned}[b]
 \mathbf R_{mn}(\mathbf k_f,\mathbf k_i)
 ={}&
 \boldsymbol{\mathcal A}_m(\mathbf k_f)
 -\boldsymbol{\mathcal A}_n(\mathbf k_i)
 \\
 &-
 \bigl(\nabla_{\mathbf k_f}+\nabla_{\mathbf k_i}\bigr)
 \arg T_{mn}(\mathbf k_f,\mathbf k_i).
\end{aligned}
\label{eq:sv-result}
\end{equation}
The two derivatives capture the phase dependence on both endpoints of the transition. Together with the Berry-connection difference they give a gauge-invariant generalized $T$-matrix shift vector.

\subsubsection{Symmetry Analysis of Generalized Shift Vector}
We start from the generalized shift vector $\mathbf R_{\alpha\beta} =\boldsymbol{\mathcal A}_\alpha-\boldsymbol{\mathcal A}_\beta-(\boldsymbol{\partial}_\alpha+\boldsymbol{\partial}_\beta)\operatorname{arg}T_{\alpha\beta}$, where $\alpha=(n,\mathbf k)$ and $\beta=(m,\mathbf k')$. . Here $\boldsymbol{\mathcal A}_\alpha=i\langle u_{n\mathbf k}| \nabla_{\mathbf k}u_{n\mathbf k}\rangle$, $\boldsymbol{\mathcal A}_\beta=i\langle u_{m\mathbf k'}|\nabla_{\mathbf k'}u_{m\mathbf k'}\rangle$,$\boldsymbol{\partial}_\alpha=\nabla_{\mathbf k}$, and $\boldsymbol{\partial}_\beta=\nabla_{\mathbf k'}$. The two derivatives act on the respective momentum arguments of the transition amplitude.

We define the exchange-symmetric and antisymmetric shift vectors as $\mathbf R_{\alpha\beta}^{p,S/A}=(\mathbf R_{\alpha\beta}^{p}\pm\mathbf R_{\beta\alpha}^{p})/2$, where $p=\linpol,\mathrm{LH},\mathrm{RH}$. We work locally in symmetry-related Bloch gauges. Under $\mathcal T$ symmetry, the partner states $|\bar\alpha\rangle=\Theta_{\mathcal T}|\alpha\rangle$ and $|\bar\beta\rangle=\Theta_{\mathcal T}|\beta\rangle$ have reversed momenta. Using
$|u_{\bar n,-\mathbf k}\rangle =\Theta_{\mathcal T}|u_{n,\mathbf k}\rangle$ and anti-unitarity gives
\begin{equation}
\begin{aligned}[b]
 \boldsymbol{\mathcal A}_{\bar\alpha}=i\langle u_{\bar n,-\mathbf k}|
 \nabla_{-\mathbf k}u_{\bar n,-\mathbf k}\rangle=-i\langle\nabla_{\mathbf k}u_{n,\mathbf k}
 |u_{n,\mathbf k}\rangle
 =\boldsymbol{\mathcal A}_\alpha.
\end{aligned}
\label{eq:sv-T-berry}
\end{equation}
The Berry connection is therefore unchanged in this gauge, while momentum reversal gives $\boldsymbol{\partial}_{\bar\alpha}=-\boldsymbol{\partial}_\alpha$. The same relations hold for $\beta$.

Using the amplitude identities $T^\linpol_{\bar\alpha\bar\beta}=T^\linpol_{\beta\alpha}$, $T^{\mathrm{LH}}_{\bar\alpha\bar\beta}=T^{\mathrm{RH}}_{\beta\alpha}$, and $T^{\mathrm{RH}}_{\bar\alpha\bar\beta}=T^{\mathrm{LH}}_{\beta\alpha}$ \cite{SakuraiNapolitano2021}, with reciprocal energy-transfer process, we obtain
\begin{align}
 \mathbf R^\linpol_{\bar\alpha\bar\beta}=-\mathbf R^\linpol_{\beta\alpha},
 \quad\mathbf R^{\mathrm{LH/RH}}_{\bar\alpha\bar\beta}=-\mathbf R^{\mathrm{RH/LH}}_{\beta\alpha}.
 \label{eq:sv-T-reversal}
\end{align}
Thus $\mathcal T$ relates a transition to its reverse and exchanges the two circular polarizations. Taking the exchange-symmetric and antisymmetric parts, we obtain Eq.~\eqref{eq.RTsymmetry} in the main text given by
\begin{equation}
 \mathbf R^{\linpol,S/A}_{\bar\alpha\bar\beta}
 =\mp\mathbf R^{\linpol,S/A}_{\alpha\beta},\quad\mathbf R^{\circpol,S/A}_{\bar\alpha\bar\beta}
 =\pm\mathbf R^{\circpol,S/A}_{\alpha\beta}.
 \label{eq:sv-T-parity}
\end{equation}
For linear polarization, the symmetric shift vector is $\mathcal T$ odd, while the antisymmetric part is $\mathcal T$ even. For the helicity-odd combination, exchanging the two helicities supplies an additional minus sign, reversing the linear-polarization parities.

Under $\mathcal{PT}$ symmetry, the partner states $|\tilde\alpha\rangle
=\Theta_{\mathcal{PT}}|\alpha\rangle$ and $|\tilde\beta\rangle =\Theta_{\mathcal{PT}}|\beta\rangle$ retain their original momenta. Using
$|u_{\tilde n,\mathbf k}\rangle =\Theta_{\mathcal{PT}}|u_{n,\mathbf k}\rangle$ gives
\begin{equation}
\begin{aligned}[b]
 \boldsymbol{\mathcal A}_{\tilde\alpha}
 =i\langle u_{\tilde n,\mathbf k}|
 \nabla_{\mathbf k}u_{\tilde n,\mathbf k}\rangle=i\langle\nabla_{\mathbf k}u_{n,\mathbf k}
 |u_{n,\mathbf k}\rangle
 =-\boldsymbol{\mathcal A}_\alpha.
\end{aligned}
\label{eq:sv-PT-berry}
\end{equation}
The Berry connection changes sign, whereas $\boldsymbol{\partial}_{\tilde\alpha}=\boldsymbol{\partial}_\alpha$ because the momentum is unchanged. The same relations hold for $\beta$.

Combining these results with $T^\linpol_{\tilde\alpha\tilde\beta}=-T^\linpol_{\beta\alpha}$, $T^{\mathrm{LH}}_{\tilde\alpha\tilde\beta}=-T^{\mathrm{RH}}_{\beta\alpha}$, and
$T^{\mathrm{RH}}_{\tilde\alpha\tilde\beta}=-T^{\mathrm{LH}}_{\beta\alpha}$ \cite{SakuraiNapolitano2021}, we obtain
\begin{align}
 \mathbf R^\linpol_{\tilde\alpha\tilde\beta}
 =\mathbf R^\linpol_{\beta\alpha},\quad
 \mathbf R^{\mathrm{LH/RH}}_{\tilde\alpha\tilde\beta}
 =\mathbf R^{\mathrm{RH/LH}}_{\beta\alpha}.
 \label{eq:sv-PT-reversal}
\end{align}
The reversal relations carry the opposite overall sign
from their $\mathcal T$ counterparts. 

By the same procedure, taking the exchange-symmetric and antisymmetric parts, we obtain Eq.~\eqref{eq.RPTsymmetry} in the main text yielding
\begin{equation}
 \mathbf R^{\linpol,S/A}_{\tilde\alpha\tilde\beta}
 =\pm\mathbf R^{\linpol,S/A}_{\alpha\beta},\quad \mathbf R^{\circpol,S/A}_{\tilde\alpha\tilde\beta}
 =\mp\mathbf R^{\circpol,S/A}_{\alpha\beta}.
 \label{eq:sv-PT-parity}
\end{equation}
For linear polarization, the symmetric shift vector is $\mathcal{PT}$ even, while the antisymmetric part is $\mathcal{PT}$ odd. In contrast, the helicity exchange again reverses these parities. Throughout, the upper and lower signs correspond
to the symmetric and antisymmetric parts, respectively.

\subsection{Non-hermitian Quantum Geometry}
\label{sec:transition-geometry}
In this section we formulate non-Hermitian quantum geometry on a transition vector bundle over the product base manifold of initial and final momentum spaces. Using the Hilbert--Schmidt inner product and the induced covariant derivative \cite{KobayashiNomizu1996} , we construct the non-Hermitian quantum metric and connection and further introduce the symmetric and skew quantum geometry.
\subsubsection{Non-Hermitian Transition Vector Bundle.}
For an $N$-band system, the Bloch eigenstates $\{|u_n(k)\rangle\}_{n=1}^N$ span the Hilbert space $\mathcal H(k)$. The complex-linear maps from $\mathcal H(k)$ to $\mathcal H(k')$ forms the fibers of the operator bundle $\mathcal{E}_{\mathbf O}$ over the base manifold $\boldsymbol{\mathcal M}=K_i\times K_f$, with projection $\pi:\mathcal E_{\mathbf O}\to\boldsymbol{\mathcal M}$ \cite{KobayashiNomizu1996}. For disorder-assisted optical transitions, $\sum_{m,n}T_{mn}=-e\sum_{m,n}E_a(\omega)t^a_{mn}(k',k)$, each polarization component defines an operator $\sum_{m,n}t^a_{mn}(k',k)|u_m(k')\rangle\langle u_n(k)|$, $a=1,\ldots,d$, which is a section of $\mathcal{E}_{\mathbf O}$.

Essentially, the product base manifold extends the single-momentum setting of Ref.~\cite{AhnEtAl2022}, with tangent space and general tangent vector given by
\begin{equation}
\begin{aligned}
 T_x\boldsymbol{\mathcal M}
 \simeq T_kK_i\oplus T_{k'}K_f,\quad
 X=v^i\partial_{k^i}+w^j\partial_{k'^j},
\end{aligned}
\label{eq:product-tangent-vector}
\end{equation}
where $x=(k,k')$ and $i,j=1,\ldots,d$. 

To describe the resonant optical transitions from states $|u_n(k)\rangle $ to $|u_m(k')\rangle$, we project the field-free optical transition operators onto $|u_m(k')\rangle\langle u_n(k)|$ which gives the transition sections
\begin{equation}
 \hat e^a_{mn}(x)
 =t^a_{mn}(x)|u_m(k')\rangle\langle u_n(k)|,
 \quad a=1,\ldots,d,
 \label{eq:transition-section}
\end{equation}
where $x=(k,k')$ and $t^a_{mn}(x)\equiv t^a_{mk',nk}(\omega)$ is the field-free optical transition amplitude. This construction underlies the geometric origin of Eq.~\eqref{eq:transition-section'} in the main text. For a fixed pair of bands $n,m$, the complex vector spaces spanned fiberwise by $\{\hat e^a_{mn}(x)\}_a$ form a vector sub-bundle $\mathcal E_{\mathbf{NHQ}}$ of $\mathcal E_{\mathbf O}$. Resonant optical transitions between these bands probe the geometry of this sub-bundle, which inherits an induced geometric structure from the complex operator bundle $\mathcal E_{\mathbf O}$~\cite{KobayashiNomizu1996}. 

In the following, we will demonstrate how the induced geometry structure in NHQ-bundle could be turned into the Non-hermitian quantum metric, connection and how do these geometry structures relate to our Hermitian and non-Hermitian part of injection and shift photocurrents.

\subsubsection{Quantum Metric}
In quantum mechanics, each fiber in the operator bundle $\operatorname{\mathcal E}_\mathbf O$ is naturally equipped with the Hilbert--Schmidt inner product,
\begin{equation}
 \begin{aligned}
 h(S_1,S_2)=\operatorname{Tr}
 (S_1^\dagger S_2),\quad
 h(S_2,S_1)=h(S_1,S_2)^*.
 \end{aligned}
 \label{eq:hs-inner-product}
\end{equation}
which forms a complex Riemannian Hermitian metric structure. Since NHQ-bundle is a sub-vector bundle of $\operatorname{\mathcal E}_\mathbf O$. The pullback construction \cite{KobayashiNomizu1996} induces the Hilbert--Schmidt tensor on the NHQ-bundle:
\begin{equation}
 Q_{mn}^{ab}
 \equiv h(\hat e_{mn}^b,\hat e_{mn}^a)
 =(t_{mn}^b)^*t_{mn}^a,
 \label{eq:transition-metric}
\end{equation}
which reproduces Eq.~\eqref{eq:transition-quantum-metric} in the main text. This tensor satisfies $Q_{mn}^{ab}=(Q_{mn}^{ba})^*$ and is positive semidefinite. The transition rate from $n\to m$ is therefore
\begin{equation}
\begin{aligned}[b]
 W_{n\to m}
 &=\frac{2\pi e^2}{\hbar}\sum_{\pm \omega}
 \Delta_\omega|T_{mn}(\omega)|^2\\
 &=\frac{2\pi e^2}{\hbar}\sum_{\pm\omega,a,b}
 \Delta_\omega E^a(\omega)E^b(-\omega)Q_{mn}^{ab},
\end{aligned}
\label{eq:transition-rate-metric}
\end{equation}
where $\Delta_\omega=\delta(\epsilon_{mn}-\hbar\omega)$ and $\epsilon_{mn}=\epsilon_m-\epsilon_n$. Momentum arguments are suppressed, and reverse quantities below are understood to belong to the reciprocal momentum and photon processes.

Following Ref.~\cite{McGinleyCooper2020}, we write the field-free transition operator as $\hat t^a=\hat{\mathcal X}^a+i\hat{\mathcal Y}^a$, where $\hat{\mathcal X}^a=[\hat t^a+(\hat t^a)^\dagger]/2$ and $\hat{\mathcal Y}^a=[\hat t^a-(\hat t^a)^\dagger]/(2i)$ are Hermitian. The quantum metric then separates into $[Q_{\rm S}]_{mn}^{ab}=\mathcal X_{nm}^b\mathcal X_{mn}^a+\mathcal Y_{nm}^b\mathcal Y_{mn}^a$ and $[Q_{\rm A}]_{mn}^{ab}=i(\mathcal X_{nm}^b\mathcal Y_{mn}^a-\mathcal Y_{nm}^b\mathcal X_{mn}^a)$. These parts are respectively even and odd under exchange of $(n,\mathbf k)$ and $(m,\mathbf k')$ combined with complex conjugation. Thus,
\begin{equation}
 [Q_{\rm S,A}]_{mn}^{ab}
 =\frac12\bigl[Q_{mn}^{ab}\pm(Q_{nm}^{ab})^*\bigr],
 \label{eq:process-metric-decomposition}
\end{equation}
which reproduces Eq.~\eqref{eq.QMSplit} in the main text. Geometrically, $[Q_{\rm S}]_{mn}^{ab}$ is Hermitian and positive semidefinite. The skew metric $[Q_{\rm A}]_{mn}^{ab}$ is also Hermitian in the polarization indices, but need not be positive semidefinite, which renders $[Q_{\rm A}]_{mn}^{ab}$ to be a non-Hermitian quantum metric~\cite{Bognar1974IndefiniteInnerProduct}. Its origin in the non-Hermiticity of the transition operator, and its vanishing in the Hermitian limit, characterize the non-Hermitian quantum geometry discussed here.

\subsubsection{Connection}
Using the same projection construction that gives rise to the NHQ quantum metrics, we define the corresponding metric-compatible connections on the NHQ-bundle. For a fixed transition process from $|u_n(k)\rangle\to|u_m(k')\rangle$, orthogonal projection onto the complex line spanned by $B_{mn}=|u_m(k')\rangle\langle u_n(k)|$ gives the basis connection \cite{KobayashiNomizu1996}
\begin{equation}
\begin{aligned}
 \omega_{mn}
 &=h(B_{mn},dB_{mn})
 \\
 &=\langle u_m(k')|du_m(k')\rangle
   +\langle du_n(k)|u_n(k)\rangle
 \\
 &=-i\left[A_i^m(k')dk'^i-A_i^n(k)dk^i\right]
 \equiv-i\mathcal A.
\end{aligned}
\label{eq:ambient-frame-derivative}
\end{equation}
Consequently, the projected connection on $\boldsymbol S=sB_{mn}$ reads
\begin{equation}
 \nabla_X\boldsymbol S
 =B_{mn}\left[X(s)-i\mathcal A(X)s\right]
 =B_{mn}\mathcal D_Xs,
 \label{eq:full-product-covariant-derivative}
\end{equation}
where the covariant derivative $\mathcal D=d-i\mathcal A$ \cite{KobayashiNomizu1996} acts on the coefficient, while $\nabla$ acts on the operator section, including its
moving frame.

We now consider the particular tangent vector $\Xi_c=\partial_{k^c}+\partial_{k'^c}$. For the transition section $\hat e_{mn}^a=t_{mn}^aB_{mn}$, where $B_{mn}=|u_m(\mathbf k')\rangle\langle u_n(\mathbf k)|$, the quantum connection evaluated along $\Xi_c$ is
\begin{equation}
\begin{aligned}[b]
 C_{mn}^{acb}
 &\equiv h(\hat e_{mn}^b,\nabla_{\Xi_c}\hat e_{mn}^a)\\
 &=(t_{mn}^b)^*
 \bigl[\partial_{k^c}+\partial_{k'^c}
 +iA_c^n(\mathbf k)-iA_c^m(\mathbf k')\bigr]
 t_{mn}^a\\
 &=(t_{mn}^b)^*\mathcal D_{\Xi_c}^{mn}t_{mn}^a,
\end{aligned}
\label{eq:transition-connection}
\end{equation}
which reproduces Eq.~\eqref{eq:transition-connection-fieldfree} in the main text. The Berry-connection terms account for the moving initial and final Bloch frames. Their contributions cancel when differentiating $Q_{mn}^{ab}=(t_{mn}^b)^*t_{mn}^a$, giving the metric-compatibility identity $\Xi_cQ_{mn}^{ab}=C_{mn}^{acb}+(C_{mn}^{bca})^*$.

We can similarly investigate the quantum connections compatible with the symmetric and skew quantum metrics. Substituting $t_{mn}^a=\mathcal X_{mn}^a+i\mathcal Y_{mn}^a$ into Eq.~\eqref{eq:transition-connection} gives $C_{mn}^{acb}=\operatorname{Tr}\{(\mathcal X_{nm}^b-i\mathcal Y_{nm}^b)B_{mn}^\dagger\nabla_{\Xi_c}[(\mathcal X_{mn}^a+i\mathcal Y_{mn}^a)B_{mn}]\}$. Expanding this expression yields
\begin{align}
 [C_{\rm S}]_{mn}^{acb}
 &=
 \mathcal X_{nm}^b\mathcal D_{\Xi_c}^{mn}\mathcal X_{mn}^a
 +\mathcal Y_{nm}^b\mathcal D_{\Xi_c}^{mn}\mathcal Y_{mn}^a,
 \label{eq:symmetric-connection-expansion}\\
 [C_{\rm A}]_{mn}^{acb}
 &=
 i\left(
 \mathcal X_{nm}^b\mathcal D_{\Xi_c}^{mn}\mathcal Y_{mn}^a
 -\mathcal Y_{nm}^b\mathcal D_{\Xi_c}^{mn}\mathcal X_{mn}^a
 \right).
 \label{eq:skew-connection-expansion}
\end{align}
The symmetric component contains the $\mathcal X$--$\mathcal X$ and $\mathcal Y$--$\mathcal Y$ terms, while the skew component contains their mixed terms and vanishes in the Hermitian limit.

Equivalently, these components are the even and odd parts under exchange of the initial and final states followed by complex conjugation:
\begin{equation}
 [C_{\rm S,A}]_{mn}^{acb}
 =\frac12\left[
 C_{mn}^{acb}\pm(C_{nm}^{acb})^*
 \right],
 \label{eq:transition-connection-decomposition}
\end{equation}
which reproduces Eq.~\eqref{SAconnection} in the main text. Here $C_{nm}^{acb}$ is evaluated at the exchanged momenta, in the reciprocal photon process, and along the correspondingly exchanged tangent direction. The associated derivatives satisfy $(\mathcal D_{\Xi_c}^{nm}s)^*=\mathcal D_{\Xi_c}^{mn}s^*$, ensuring that this projection reproduces the explicit expansions above.

Applying the same even and odd projections to the metric-compatibility identity \cite{KobayashiNomizu1996} gives
\begin{align}
 \Xi_c[Q_{\rm S}]_{mn}^{ab}
 &=[C_{\rm S}]_{mn}^{acb}
 +([C_{\rm S}]_{mn}^{bca})^*,
 \label{eq:symmetric-metric-compatibility}\\
 \Xi_c[Q_{\rm A}]_{mn}^{ab}
 &=[C_{\rm A}]_{mn}^{acb}
 +([C_{\rm A}]_{mn}^{bca})^*.
 \label{eq:skew-metric-compatibility}
\end{align}
Thus the symmetric and skew quantum connections satisfy the compatibility identities for their respective quantum metrics.
\bibliography{NHPC_references}
\end{document}